\documentclass[fleqn,10pt]{wlscirep}

\usepackage[utf8]{inputenc}
\usepackage[T1]{fontenc}

\usepackage{multirow}	
\usepackage{Tabbing}	
\usepackage{relsize}	
\usepackage{textcomp}	
\usepackage{soul}	

\usepackage{amsmath}
\usepackage{amssymb}
\usepackage{bm}		

\usepackage{hyperref}
\usepackage{url}
\usepackage[capitalise]{cleveref}

\usepackage{dcolumn}	
\usepackage{longtable}	
\usepackage{booktabs}	

\usepackage{graphicx}
\usepackage{epstopdf}	
\usepackage{rotating}	
\usepackage{float}	
\usepackage{verbatim} 

\usepackage{color}

\newcommand{\beginsupplement}{%
        \setcounter{table}{0}
        \renewcommand{\thetable}{S\arabic{table}}%
        \setcounter{figure}{0}
        \renewcommand{\thefigure}{S\arabic{figure}}%
        \setcounter{equation}{0}
        \renewcommand{\theequation}{S\arabic{equation}}%
        \setcounter{page}{1}
        \renewcommand{\thepage}{S\arabic{page}}%
        \renewcommand{\numpages}{\pageref{PAG:endSupp}}
     }

\DeclareMathOperator{\sgn}{sgn}

\newcommand{\bra}[1]{\ensuremath{\left\langle#1\right|}}
\newcommand{\ket}[1]{\ensuremath{\left|#1\right\rangle}}
\newcommand{\average}[1]{\ensuremath{\left\langle#1\right\rangle}}
\newcommand{\bracket}[2]{\ensuremath{\left\langle#1 \vphantom{#2}\right| \left. #2 \vphantom{#1}\right\rangle}}
\newcommand{\matrixel}[3]{\ensuremath{\left\langle #1 \vphantom{#2#3} \right| #2 \left| #3 \vphantom{#1#2} \right\rangle}}
\newcommand{\kron}[2]{\ensuremath{\delta_{#1#2}}}

\newcommand{\AN}[2]{
\ensuremath{\hat{c}^{} _{{#1}
\ifnum#2=1  \uparrow
\else
\ifnum#2=-1  \downarrow
\else
\ifnum#2=2  \sigma
\else
\ifnum#2=-2  \bar{\sigma}
\else
\ifnum#2=3  \sigma '
\fi
\fi
\fi
\fi
\fi
}}
}
\newcommand{\CR}[2]{
\ensuremath{\hat{c}^\dagger _{{#1}
\ifnum#2=1  \uparrow
\else
\ifnum#2=-1  \downarrow
\else
\ifnum#2=2  \sigma
\else
\ifnum#2=-2  \bar{\sigma}
\else
\ifnum#2=3  \sigma '
\fi
\fi
\fi
\fi
\fi
}}
}
\newcommand{\NUM}[2]{
\ensuremath{\hat{n}^{} _{{#1}
\ifnum#2=1  \uparrow
\else
\ifnum#2=-1  \downarrow
\else
\ifnum#2=2  \sigma
\else
\ifnum#2=-2  \bar{\sigma}
\else
\ifnum#2=3  \sigma '
\fi
\fi
\fi
\fi
\fi
}}
}

\newcommand{\mc}[3]{\multicolumn{#1}{#2}{#3}}

\renewcommand{\vec}[1]{\ensuremath{\mathbf{#1}}}

\renewcommand{\numpages}{\pageref{PAG:end}}

\begin{document}

\title{Dot-ring nanostructure: Rigorous analysis of many-electron effects}

\author[1,$\dagger$]{Andrzej Biborski}

\author[2,$\ddagger$]{ Andrzej P. Kądzielawa}

\author[3]{Anna Gorczyca-Goraj}

\author[3]{Elżbieta Zipper}

\author[3,$\S$]{Maciej M. Ma\'{s}ka}

\author[2,1,*]{J\'{o}zef Spa\l{}ek}

\affil[1]{
Akademickie Centrum Materiałów i Nanotechnologii,
AGH Akademia Górniczo-Hutnicza,
Al. Mickiewicza 30,
PL-30-059 Kraków, Poland
}
\affil[2]{                                                                 
Instytut Fizyki im. Mariana Smoluchowskiego, Uniwersytet Jagielloński, ul. Łojasiewicza 11, PL-30-348 Krak\'{o}w, Poland
}
\affil[3]{
Instytut Fizyki, Uniwersytet Śl\k{a}ski, ul. Uniwersytecka 4, PL-40-007 Katowice, Poland
}

\affil[$\dagger$]{biborski@agh.edu.pl}
\affil[$\ddagger$]{kadzielawa@th.if.uj.edu.pl}
\affil[$\S$]{maciej.maska@phys.us.edu.pl}
\affil[*]{corresponding author: ufspalek@if.uj.edu.pl}

\date{\today}

\begin{abstract}
We discuss the quantum dot-ring nanostructure (DRN) as canonical example of a nanosystem, for which the~interelectronic interactions can be evaluated exactly.
The system has been selected due to its tunability, i.e., its electron wave functions 
can be modified much easier than in, e.g., quantum dots.
We determine many-particle 
states for $N_e=2$ and $3$ electrons and calculate the 3- and 4-state
interaction parameters, and discuss their importance.
For that purpose, we combine the first-
and second-quantization schemes and hence are able to single out the component single-particle contributions
to the resultant many-particle state. The method provides both the ground- and the first-excited-state energies,
as the exact diagonalization of the many-particle Hamiltonian is carried out.
DRN provides one of the few examples for which one can determine
theoretically \textbf{all} interaction microscopic parameters to a high accuracy.
Thus the evolution of the single-particle vs. many-particle contributions to each state and its energy can be determined
and tested with the increasing system size. In this manner, we contribute to the wave-function engineering with the~interactions included for those few-electron systems.
\end{abstract}

\maketitle

\section*{Introduction and Motivation}

Few-electron systems represent a very interesting topic in quantum nanophysics \cite{Kouwenhoven}, as their studies are at the forefront of nanoelectronic
applications \cite{Ihn}, e.g., as single-electron transistors \cite{Kastner,Hanson2008} or other devices 
\cite{Hanson2007,Stopa2002,Scheibner2008}.

\begin{figure}[h!]
 \centering
 \includegraphics[width=.7\linewidth]{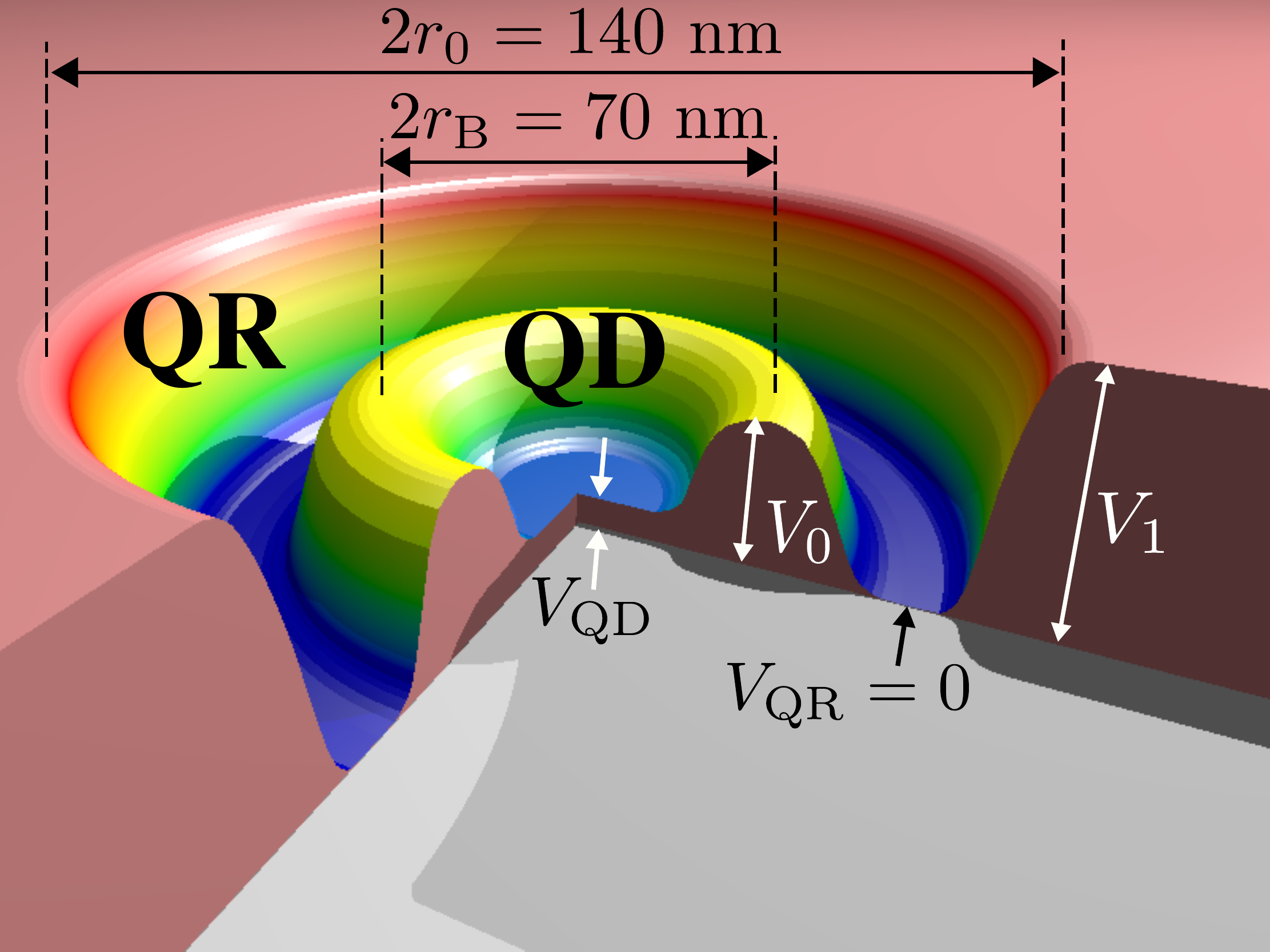} \\
 \includegraphics[width=.7\linewidth]{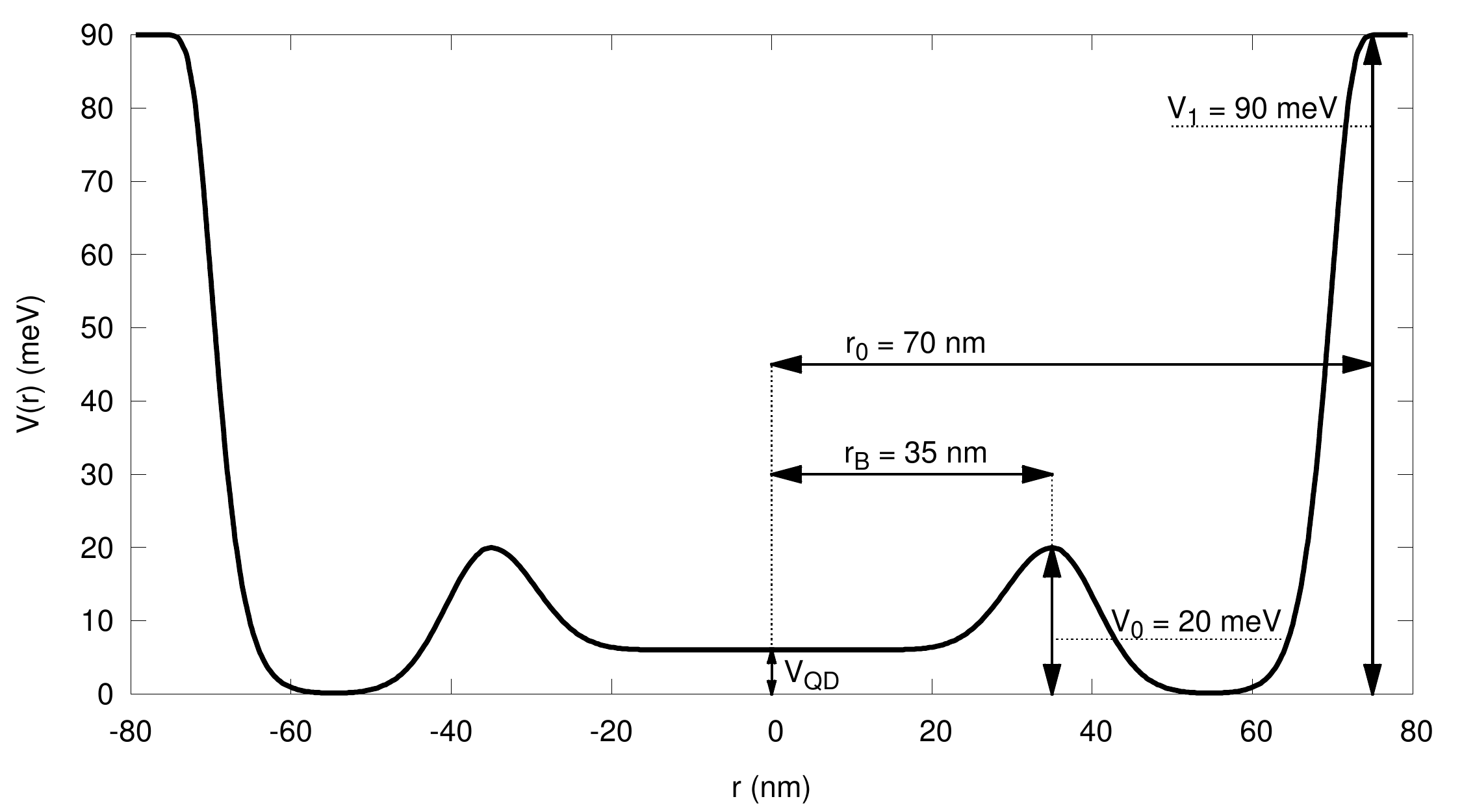} 
 \caption{Schematic representation of quantum-dot (QD) -- ring (QR) structure (top) into DRN and the shape of the actual single-particle
 potential energy (bottom), with the corresponding values taken in the analysis.}
  \label{Fig:drn_str}
\end{figure}

Recently, the basic issue of the wave-function manipulation has been raised on the example of quantum-dot--ring nanostructure, DRN \cite{Zipper1} (cf. Fig.~\ref{Fig:drn_str}).
Explicitly, the transition between single-particle states with the dominant quantum dot (QD) or ring (QR) contributions may lead
to interesting optical absorption and transport properties \cite{Zipper1,Kurpas1,Zeng}. In this context, an interesting question
arises as to what happens if the multi-electron states are involved (e.g., with the number of particles $N_e = 2, \ 3,\ \dots$).
Such problem has been addressed earlier \cite{Szafran}, where the spin and the charge switching in the applied magnetic
field has been analyzed in detail. The results demonstrate that such model system can reflect the situation encountered
in experimentally constructed devices of DRN type \cite{Somaschini21,Somaschini22,somaschini2010}.

In this paper our aim is somewhat more fundamental. Namely, we include in a rigorous manner the interelectronic interactions
for a preselected (finite) basis of single-particle states, appropriate for the system geometry. The experimentally
controlled parameter is the gate electrostatic potential $V_\mathrm{QD}$ of the quantum dot (QD) relative to that of the ring (QR).
We determine next the system energy for $N_e=2$ and $3$ electrons, as well as the many-particle wave function. This, in turn,
allows us to construct the particle-density profiles and in particular, the partial contribution of the component single-particle-state products to the many-particle ground- and the first-exited-states. Such a decomposition into the single-particle product components
is possible in the method we use, in which we combine the first- and second-quantization schemes of determining
the many-particle state. In essence, the many-particle Hamiltonian in the occupation number representation (Fock space)
is diagonalized starting from the preselected set of single-particle states in the Hilbert space
providing the scenario for possible multiple-particle occupation configurations.
For the original presentation and application of the method to various nanoscopic systems see \cite{Spalek1, Kadzielawa1, Biborski, Spalek2}.
Explicitly, we predetermine the lowest 10 single-particle states $\{ \psi_{i\sigma}(\vec{r}) \}$
for given shape of DRN potential. Those single-particle states (obtained numerically for given topology of the device) are used as an input to define the field
operators ($\hat{\Psi}_{\sigma}^{\phantom{\dagger}}(\vec{r})$ and its Hermitian conjugate counterpart $\hat{\Psi}_{\sigma}^{\dagger}(\vec{r})$, respectively)
by the prescription
\begin{align}
 \label{eq:field_operator}
 \hat{\Psi}_{\sigma}^{\phantom{\dagger}}(\vec{r}) = \sum_{i=1;\sigma=\pm 1}^{M} \psi_{i\sigma} (\vec{r}) \AN{i}{2} ; \ \ \ \hat{\Psi}_{\sigma}^{\dagger}(\vec{r}) \equiv \left( \hat{\Psi}_{\sigma}^{\phantom{\dagger}}(\vec{r}) \right) ^\dagger = \sum_{i=1;\sigma=\pm 1}^{M} \psi_{i\sigma}^* (\vec{r}) \CR{i}{2},
\end{align}
where $\AN{i}{2}$ (and $\CR{i}{2}$) are the annihilation (creation) operators of particle in the single-particle state 
$\ket{\psi_{i\sigma} (\vec{r})}$. Note that the number $M$ of states included in definition of the field operator is 
selected in such a manner that any further enrichment of the single-particle basis $\left\{ \psi_{i\sigma} (\vec{r}) \right\}$
does not change quantitatively the characteristics of the ground and the first excited states. Here, it is sufficient to take $M=10$.
In effect, no problem connected with the basis incompleteness should arise. This formal point will also be discussed
\emph{a posteriori}.

The next step is to define many-particle Hamiltonian in the second-quantization language in a standard manner (cf. e.g.
\cite{FetterWalecka}) which we diagonalize in a rigorous manner. This last step allows for determination of the system
global characteristics such as the total system energy, the multiparticle wave function, the particle density profile $n(\vec{r})$, the total
spin, and the energies of the transition between the states, e.g., the spin singlet--triplet transition for $N_e=2$,
etc. What is equally important, we calculate \textbf{all} the microscopic interaction parameters $V_{ijkl}$, including
the 3-- (e.g., $V_{ijki}$) and 4-state parameters $V_{ijkl}$, i.e., those with all the indices different. In result, we can discuss explicitly the importance of those nontrivial terms, which are often neglected even in many-particle considerations \cite{Kadzielawa1,Biborski,Spalek2}.
We believe that this last result, coming from our method should be taken into consideration, as those interactions are
often non-negligible, to say the least. In any case, they should be evaluated to see their relevance, at least in model situations.

The structure of the paper is as follows. We define first the Hamiltonian and detail the method of
calculations. Next, we discuss the basic characteristics of the multiparticle states, as well as
determine the values of \textbf{all} nontrivial microscopic parameters. Finally, we determine the energy of the 
singlet-triplet transition (for $N_e=2$), as well as discuss the doublet-quadruplet transition for $N_e=3$,
which should be detectable in the microwave domain. At the end, we discuss briefly the application of our results
to determine the optical transitions and, e.g., the transport of electrons throughout such system. In Supplementary Material we display the shapes of the starting 
single-particle wave functions, provide detailed numerical values of the 3- and 
4-state interaction parameters, as well as display the detailed system characteristics for selected values of $V_\mathrm{QD}$. In particular, in Supplement D we show the first two states degeneracy which contains a chiral factor to it, depending on the number of ways the orbital currents can be arranged for given conserved total quantum numbers $S^z_{tot}$, $\vec{S}^2_{tot}$ and $L^z_{tot}$.

\section*{Problem and Method}
\label{sec:problem}
We start from the single--particle solution of the Schr\"odinger equation for the DRN system parametrized as in \cite{Kurpas1}. Therefore, 
the set of the single-particle eigenfunctions $\psi_{n,l}(r,\phi,z)$ in the cylindrical coordinates, being the solution for the one-electron DRN picture, is assumed at the start \cite{Zipper1,Kurpas1,Zeng}.
The many--particle problem in which electrons are described by the second quantizied Hamiltonian has the standard form \cite{FetterWalecka}
\begin{align}
 \label{eq:hamiltonian}
  \mathcal{H} &\equiv \sum_{\sigma} \int d^3 r \hat{\Psi}^\dagger_\sigma (\vec{r}) \mathcal{H}_1  \hat{\Psi}^{\phantom{\dagger}}_\sigma (\vec{r}) + \frac{1}{2} \sum_{\sigma \sigma '} \iint d^3 r d^3 r' \hat{\Psi}^\dagger_\sigma (\vec{r}) \hat{\Psi}^\dagger_{\sigma '} (\vec{r}') V(\vec{r}-\vec{r}')  \hat{\Psi}^{\phantom{\dagger}}_{\sigma '} (\vec{r}')  \hat{\Psi}^{\phantom{\dagger}}_\sigma (\vec{r})\\\notag
  &= \sum\limits_{ij}\sum\limits_{\sigma}t_{ij}\CR{i}{2}\AN{j}{2} +\frac{1}{2} \sum\limits_{ijkl}\sum\limits_{\sigma,\sigma'}V_{ijkl}\CR{i}{2}\CR{j}{3}\AN{l}{3}\AN{k}{2},
\end{align}
where $t_{ij} \equiv \matrixel{\psi_i}{\mathcal{H}_1}{\psi_j}$ and $V_{ijkl} \equiv \matrixel{\psi_i \psi_j}{V_{12}}{\psi_k \psi_l}$ are the microscopic parameters
which are calculated in the basis $\left\{ \psi_{i\sigma} \equiv \psi_i \chi_\sigma \right\}$. The spin--orbit interaction is neglected. In effect, the changes with respect to the corresponding
one-particle considerations \cite{Zipper1,Kurpas1} are induced solely by the interparticle interactions.
The symbols $i,j,k,l \in  \big \{ [0 \ 0], [0 \ 1], [0 \ \bar{1}], \dots [n \ l],\dots \big\}$ represent quantum number pairs referring to a single--particle solution $[n \ l]$ \cite{Kurpas1}.
One specific feature of the problem should be noted. Namely, since the single-particle wave-functions $\left\{ \psi_i (\vec{r}) \chi_\sigma \right\}$
represent the eigenfunctions of the single-particle Hamiltonian, i.e., $\mathcal{H}_1 \psi_i (\vec{r}) =  \epsilon_i \psi_i (\vec{r}) $, the first term
in \eqref{eq:hamiltonian} is explicitly diagonal, i.e., $t_{ij} = \epsilon_i \delta_{ij}$. Therefore, the diagonalization of the Hamiltonian \eqref{eq:hamiltonian}
means that such a procedure is applied to the interaction part (the second term).

To solve many-electron problem for a fixed number $N_e$ of electrons, one must proceed in two steps:
 \begin{enumerate}
  \item Compute explicitly one- and two-body microscopic parameters, $\big\{ t_{ii} \big\}$ and $\big\{ V_{ijkl}  \big\}$, respectively. \label{en:01}
  \item Diagonalize the Hamiltonian \eqref{eq:hamiltonian} in the Fock space.\label{en:02}
 \end{enumerate}
Each of these steps is discussed below. But first, we have to define the starting single-particle wave functions in the real-number domain.
 
\subsection*{Change to the real single-particle basis functions}
Eigenfunctions  $\psi_{n,l}(r,\phi,z)$  -- by their nature -- form an orthogonal and normalized single-particle basis of planar rotational symmetry \cite{Zipper1,Kurpas1},
\begin{align}
 \label{eq:orthogonality}
  \bracket{\psi_{n,l}}{\psi_{n',l'}}=\delta_{nn'}\delta_{ll'},
\end{align}
where in the cylindrical coordination system $(r,\phi,z)$ we have that
\begin{align}
 \label{eq:psi}
  \psi_{n,l} = e^{il\phi}\chi_{n}(r,z).
\end{align}
As the microscopic parameters are to be calculated numericaly (since the explicit analytical form of the single-particle wave functions is not known), it is convenient to deal with the real-space basis. 
Hence, we utilize the real representation, exploiting in fact the cylindrical geometry of problem, namely
\begin{align}
\label{real_transformation}
\varphi_{nl}(r,\phi,z) \equiv \frac{\psi_{n,|l|} + \sgn(l)\psi_{n,-|l|}}{\sqrt{2\sgn(l)}}.
\end{align}

\subsection*{Microscopic parameters computation}

\begin{figure}[h]
 \includegraphics[width=.7\linewidth]{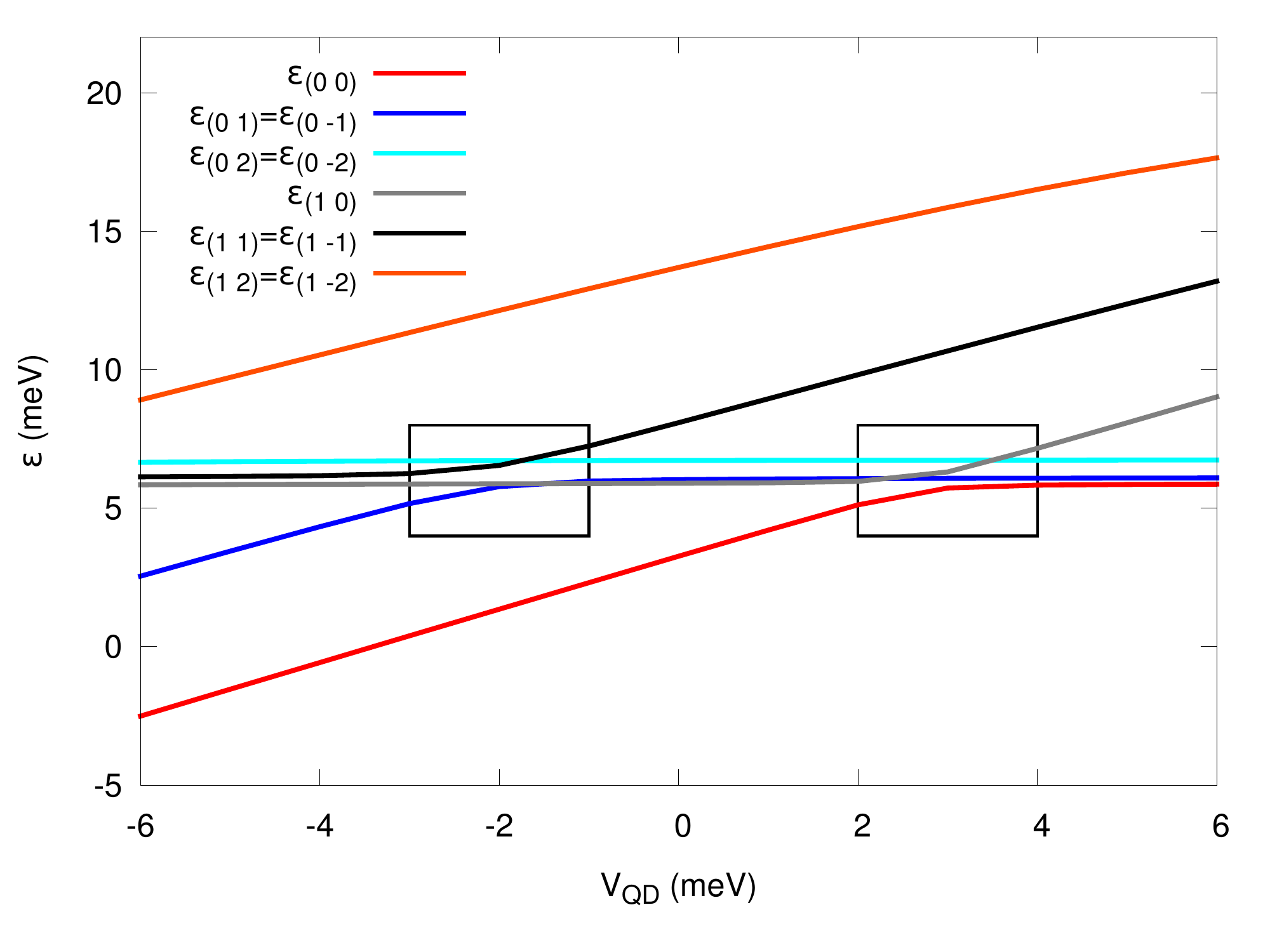} 
 \caption{Single-particle energies $\{ \epsilon_i \}$ for the first ten wave-functions versus QD potential energy $V_\mathrm{QD}$. 
 One can see that some the energies vary with the value of $V_\mathrm{QD}$, whereas the others are independent of $V_\mathrm{QD}$. States in the former group are 
 located in the QD part of the DRN and the latter states in the QR part.
 Note the two regions of the level crossing or anticrossing (framed).
 In these regimes, some of the states, with the increasing $V_\mathrm{QD}$, change over from the QD to the QR as the dominant regions (after \cite{Kurpas1}).}
  \label{Fig:sp_energy}
\end{figure}

The transformation \eqref{real_transformation} preserves both the orthogonality and the normalization of starting wave functions and can be applied to the computation
of the microscopic parameters defining  Hamiltonian \eqref{eq:hamiltonian}. Evaluation of single--particle parameters $t_{ij}$ is performed in terms of integration in the new basis, namely
 \begin{align}
 \label{eq:onbodyint}
  t_{ij} = \bra{\varphi_i(\vec{r})}\mathcal{H}_1{}\ket{\varphi_j(\vec{r})} \equiv \int d^3 r \varphi_i(\vec{r}) \mathcal{H}_1 \varphi_j(\vec{r}).
\end{align}
However, as said above, since eigenproblem of one electron is solved \cite{Kurpas1},
the eigenvalues $t_{ii} = \epsilon_i$ are known (cf. Fig.~\ref{Fig:sp_energy}).
Furthermore,  the elements $t_{ij}$ for $i\neq j$ vanish also after the basis transformation to the form \eqref{real_transformation}.
For the sake of clarity, we define $\varphi_{nl}(r,\phi,z) \equiv \varphi_{nl}$ and label $-|l| \equiv \bar{|l|}$ to write explicitly that
 \begin{align}
 \label{eq:lemma_1}
  t_{ij} &= \bra{\varphi_{nl}}\mathcal{H}_1{}\ket{\varphi_{n'\bar{l}'}} = 
  \bra{\frac{\psi_{n,|l|} + \sgn(l)\psi_{n,\bar{|l|}}}{\sqrt{2\sgn(l)}} }\mathcal{H}_1{}\ket{{\frac{\psi_{n',|l'|} + \sgn(l')\psi_{n',\bar{|l'|}}}{\sqrt{2\sgn(l')}} }} \\[5pt] \notag
  &=\frac{\bra{\psi_{n|l|}}\mathcal{H}_1\ket{\psi_{n'|l'|}} + 
  \sgn(l')\bra{\psi_{n|l|}}\mathcal{H}_1\ket{\psi_{n'\bar{|l'|}}} +  \sgn(l)\bra{\psi_{n\bar{|l|}}}\mathcal{H}_1\ket{\psi_{n'|l'|}}  +  \sgn(ll')\bra{\psi_{n\bar{|l|}}}\mathcal{H}_1\ket{\psi_{n'\bar{|l'|}}}}{2\sqrt{\sgn(ll')}}\\[5pt] \notag
  &=\frac{\kron{n}{n'}\kron{|l|}{|l'|}\epsilon_{n'|l'|} + \sgn(l')\kron{n}{n'}\kron{|l|}{\bar{|l'|}}\epsilon_{n'\bar{|l'|}} + \sgn(l)\kron{n}{n'}\kron{\bar{|l|}}{|l'|}\epsilon_{n'|l'|} + \sgn(ll')\kron{n}{n'}\kron{\bar{|l|}}{\bar{|l'|}}\epsilon_{n'\bar{|l'|}}}{2\sqrt{\sgn(ll')}}\\[5pt] \notag
  &=\frac{\kron{|l|}{|l'|}\epsilon_{n'|l'|}\big[1+\sgn(ll')\big]}{2\sqrt{\sgn(ll')}}
     =\begin{cases}
     \displaystyle \epsilon_{nl}=\epsilon_{i} &\mbox{if } l = l' \wedge n = n'  \\[5pt]
     \displaystyle 0 &\mbox{if } l\neq l' \vee n \neq n'
 \end{cases},
\end{align}
where $\sgn(l)$ is the sign function. We also utilize symmetry of the single-particle solution, i.e., $\epsilon_{nl} = \epsilon_{n\bar{l}}$.

Now, the two-body (four-state) integrals $V_{ijkl}$ are expressed as
\begin{align}
 \label{eq:twobodyint}
  V_{ijkl} = \bra{\varphi_i(\vec{r})\varphi_j(\vec{r'})}\frac{e^2}{4\pi\varepsilon_{0}\varepsilon|\vec{r}-\vec{r'}|}\ket{\varphi_k(\vec{r})\varphi_l(\vec{r'})},  
\end{align}
where $e$ is the electron charge, $\varepsilon_{0}$ is the vacuum permittivity and $\varepsilon = 12.9$ is the relative permittivity, taken here for the $GaAsIn$ system.
Their explicit determination is required for a further Hamiltonian matrix construction.
These, up to four-state integrals, are six-dimensional and therefore, standard numerical  integration techniques are not suitable for this task. Instead, the Monte-Carlo integration 
scheme has been applied.  For that aim, we use CUBA library \cite{CUBA}, selecting the \emph{suave} algorithm for the integrals calculations. The procedure is standard and the accuracy
of such integration is typically $0.005$ meV or even better.

\subsection*{Method: diagonalization of the multiparticle Hamiltonian}

We start from the occupation number representation of the multiparticle states in the Fock space in the following form
\begin{align}
\label{eq:basis_states_eg}
 \ket{\Phi} =& \underbrace{\ket{0, 1, \dots, 1}}_{\text{spin } \uparrow} \otimes \underbrace{\ket{1, 0, \dots, 1}}_{\text{spin } \downarrow} = \\\notag
	    =& \CR{2}{1} \cdots \CR{M}{1} \CR{1}{-1} \cdots \CR{M}{-1} \ket{0},
\end{align}
where $M$ is the number of states.
We find explicitly all the possible states for $N_e$ electrons and thus are able to build up Hamiltonian matrix out of \eqref{eq:hamiltonian}
by calculating all the averages $\matrixel{\Phi'}{\mathcal{H}}{\Phi}$. We diagonalize the resultant matrix using the \emph{QR decomposition} of the
Gnu Scientific Library (GSL) \cite{GSL}. The usage of Lanczos algorithm is not efficient in this case, as both the ground and the first excited
states can be highly degenerate. \emph{The QR decomposition}, as well as the GSL library, operate with relatively small matrices
(of dimension not exceeding $10^5 \times 10^5$ elements), but this is not the number of states to be reached for small number of electrons, 
even for a relatively large number of sinle--particle wave--functions included in the starting basis \eqref{eq:field_operator}.

For the purpose of these calculations we employ also our library \emph{the Quantum Metallization Tools} (QMT)  \cite{qmtURL}, proved to be efficient for similar problems \cite{Biborski}.
Explicitly, the calculations of the parameters $t_{ij}$ and $V_{ijkl}$ in \eqref{eq:hamiltonian} have been carried out with the help of the Monte-Carlo (MC) integration method described in 
\cite{CUBA}. The accuracy of their evaluation is estimated as $0.005$ meV. The validity of application of MC in the current context was tested  by means of
a numerical computation of the \emph{on--site} $1s$ electron--electron interaction for the Slater function, for which an analytical formula exists.

\section*{Results: Two- and Three-Electron States}
\label{sec:results}

\subsection*{Basic Characteristics}

\begin{figure}[h!]
 \includegraphics[width=0.7\linewidth]{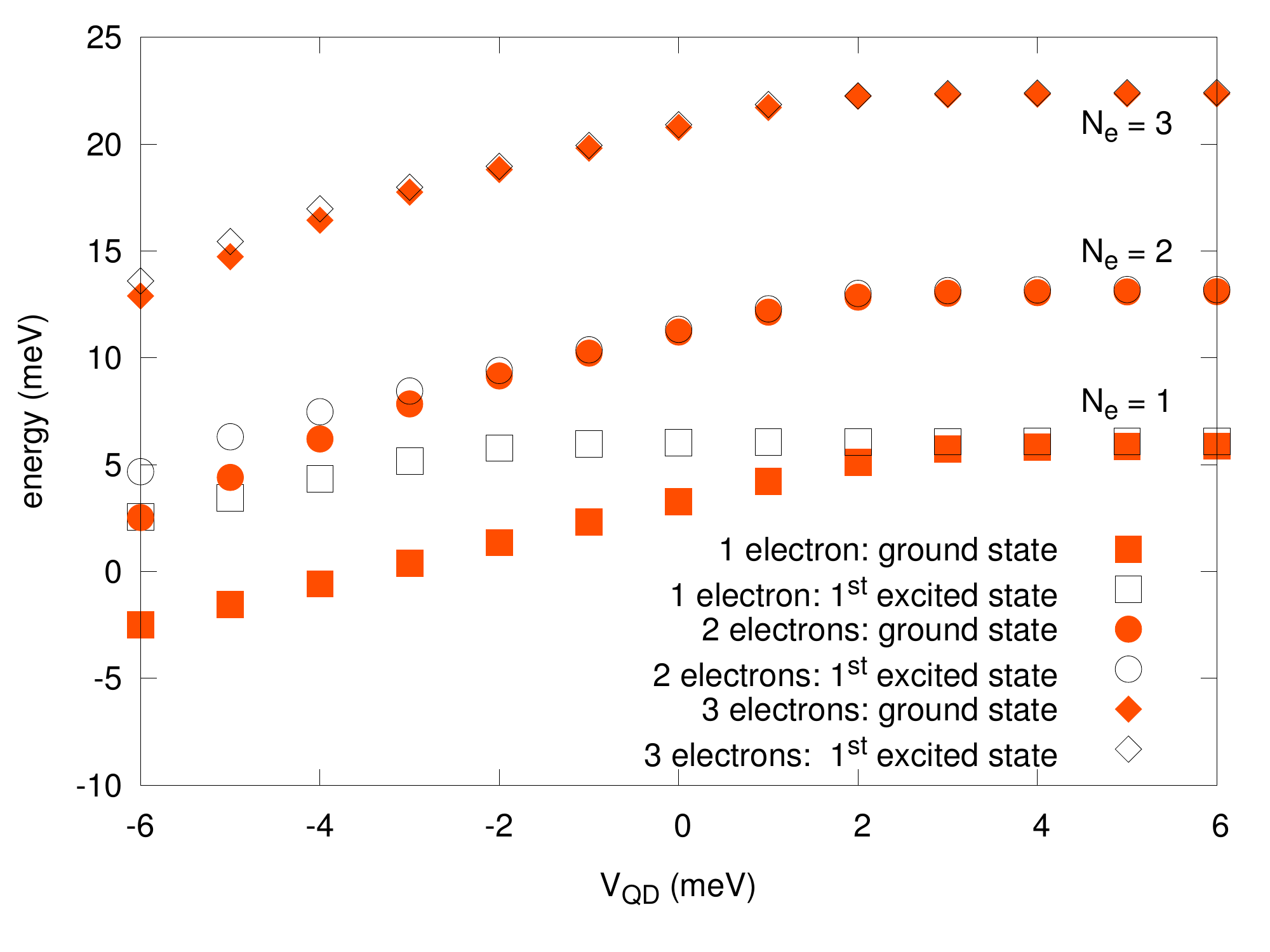} 
 \caption{Ground and first excited state energies for $N_{e}$ = $1$, $2$ and $3$ electrons in DRN versus the QD potential energy.
 The single-particle energy (bottom squares) is provided for comparison. The interelectronic interactions alter essentially the resultant energies.
 Note that roughly the energy for $N_{e}$ = $3$ increases with respect to the case $N_{e}$ = $2$ by the factor $N_{e}(N_{e} - 1)/2$.}
  \label{Fig:energy}
\end{figure} 
We are interested in calculating the system observables. In this Section we present the results for basic quantities, in this case 
the energy, and the total electronic density $n(\vec{r}) \equiv \sum_\sigma \average{\hat{\Psi}^\dagger_\sigma (\vec{r}) \hat{\Psi}^{\phantom{\dagger}}_\sigma (\vec{r}) } $ in 
the many-particle state. The states are characterized by the conserved quantities, i.e., the $z$-component $L^z$
of the angular momentum,
the total spin $\average{\vec{S}_{tot} ^2}$, and its $z$-component $\average{S^z}$.
Explicitly, in Fig.~\ref{Fig:energy} we plot the ground and excited state energies for $N_e = 1$, $2$, and $3$ (curves from bottom to top, respectively).
The energy increases substantially with each particle added to the system, as expected for the Coulomb system of charges. The single-particle part of the potential energy $\sim -|e|V_\mathrm{QD} < 0$
represents a substantial contribution for its value $\sim$ few $eV_\mathrm{QD}$, comparable to that introduced by the repulsive interaction for $N_e=2$ and $3$.

\subsubsection*{Two Electrons}

Here we present electronic density, as well as $\average{\vec{S}_{tot} ^2}$ and $\average{S_{tot}^z}$ for the ground and first excited states of DRN for 2 electrons. The ground state
is always the spin-singlet $S=0$ ($\average{\vec{S}_{tot} ^2}=0$) state, whereas the first excited state is the spin-triplet
$S=1$ ($\average{\vec{S}_{tot} ^2}=2$).

\begin{figure*}[htbp]
1)\ \raisebox{-118pt}{\includegraphics[width=0.9\textwidth]{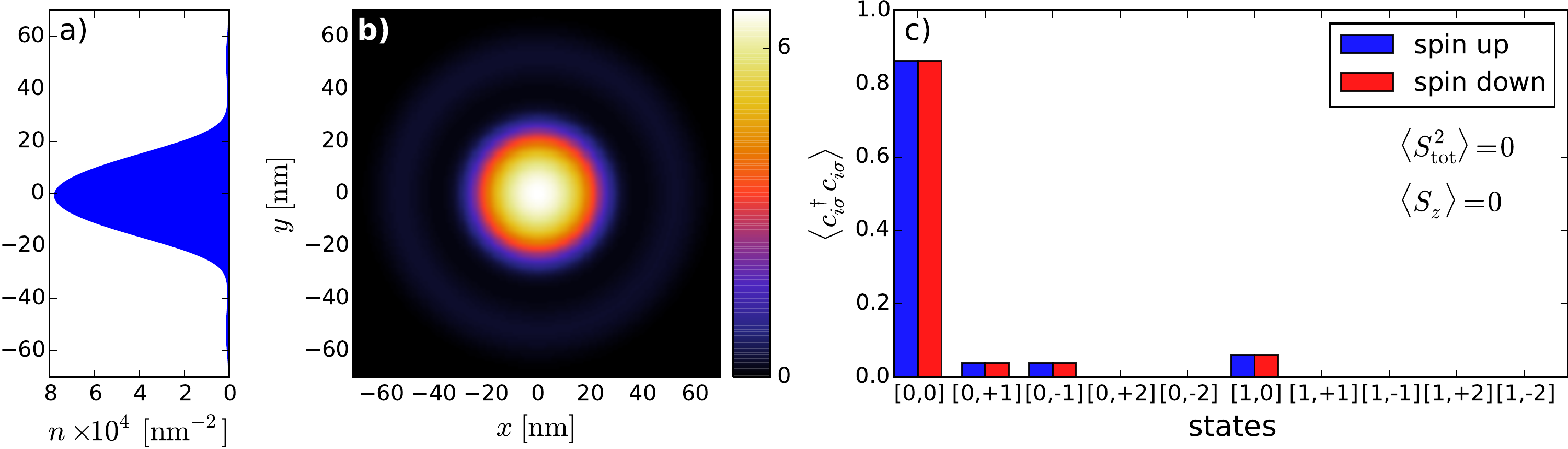}}\\
2)\ \raisebox{-118pt}{\includegraphics[width=0.9\textwidth]{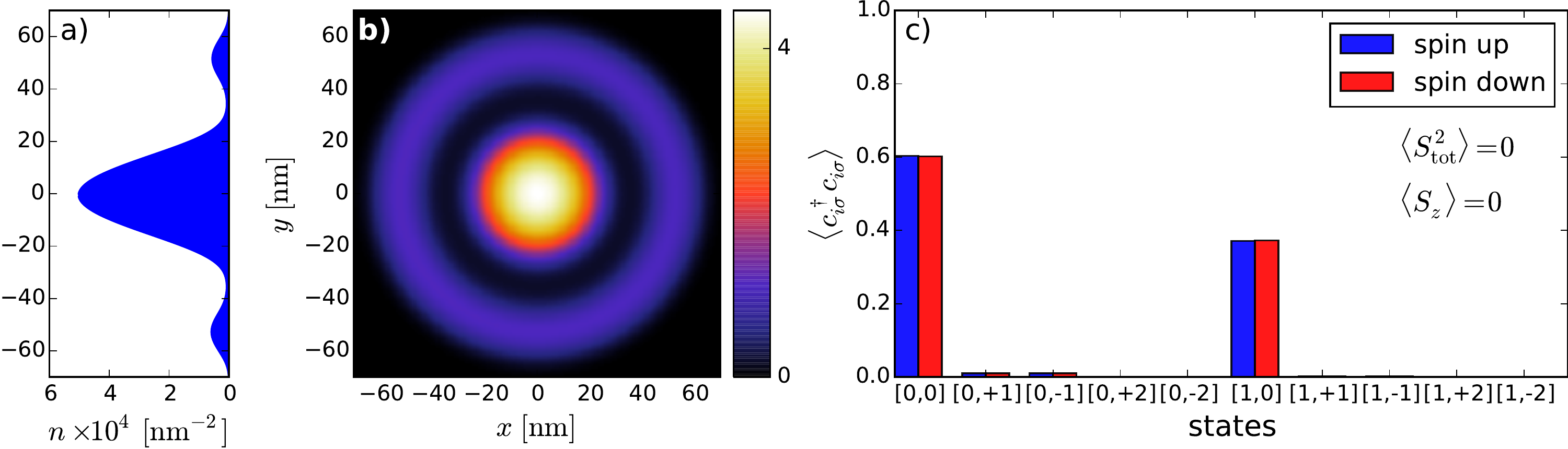}}\\
3)\ \raisebox{-118pt}{\includegraphics[width=0.9\textwidth]{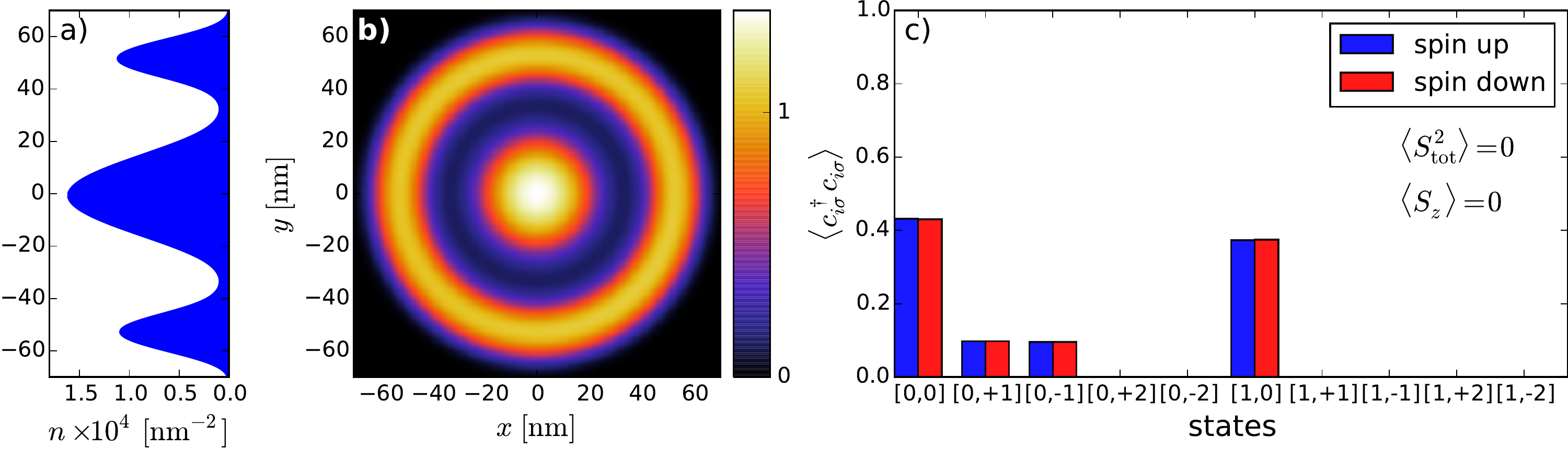}}\\
4)\ \raisebox{-118pt}{\includegraphics[width=0.9\textwidth]{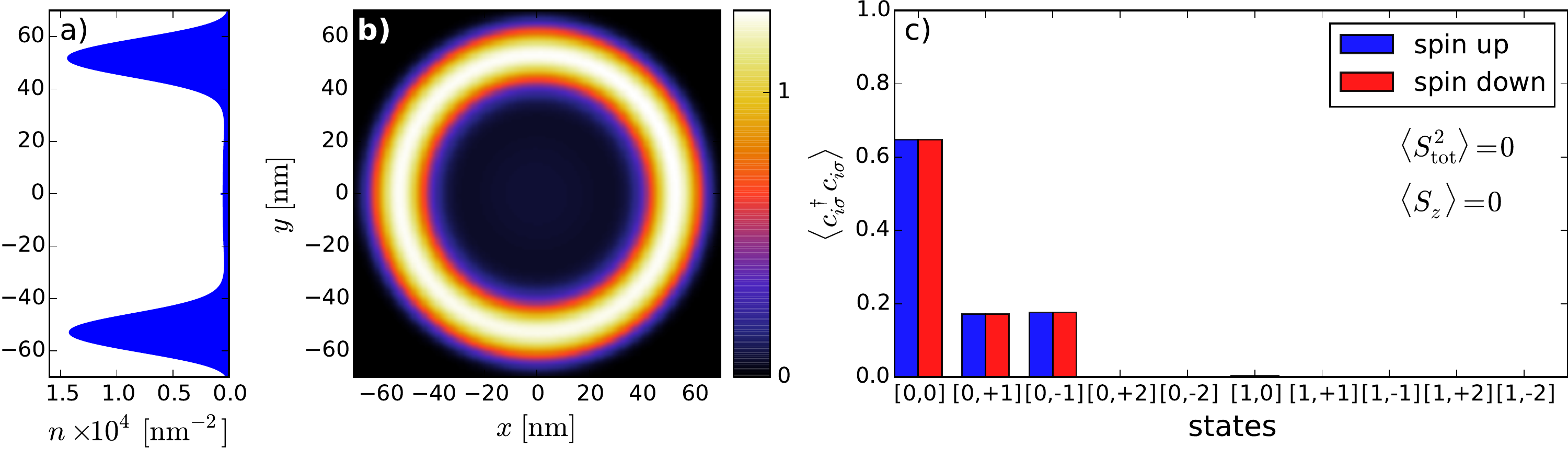}}\\
 \caption{Evolution of the electronic density profiles $n(x,y)$ [a) and b)] and the occupancy $\bar{n}_{i\sigma} \equiv \average{\CR{i}{2} \AN{i}{2}}$ of the single-particle states c) composing the many particle state
   for $N_e=2$.
   Rows correspond to $V_\mathrm{QD}$ equal to -4 meV, -2 meV, 2 meV, and 4 meV, from top to bottom respectively.
   First ten single-particle states have been taken into account to compose the resultant two-particle
 state for given $V_\mathrm{QD}$. The occupancy of the higher in energy
 single-particle states is negligible; those are shown in c) for completeness.
  \label{Fig:2DRN_m4p4}}
\end{figure*}

As can be seen in Fig. \ref{Fig:2DRN_m4p4}, with the increasing $V_\mathrm{QD}$ from $-4$ meV to $+6$ meV there is a gradual shift of dominant
part of the electron density from QD to QR. If the bottom of the central part of the confining potential is very low, the electron density is the largest within the dot part of DRN as attractive $V_\mathrm{QD}$ in this case is comparable or larger than the interaction energy. In this regime [row 1) in Fig. \ref{Fig:2DRN_m4p4}] the single particle state with $n=0$ and $l=0$ gives the main contribution to the two-particle state. When $V_\mathrm{QD}$ becomes less negative the Coulomb interaction partially ``pushes out'' the electron density towards
  the outer part of the DRN [row 2) in Fig. \ref{Fig:2DRN_m4p4}]. It is realized by increasing the contribution
  of the single particle state with $n=1$ and $l=0$ to the two-particle wave function. With further increase of $V_\mathrm{QD}$ it becomes energetically favourable to reduce the occupancy of QD, i.e., in the area where the interaction is strong due to a strong confinement in a small area. As a result, the electron density increases in QR and single-particles states with nonzero angular momenta become occupied. Finally, for $V_\mathrm{QD}=4$ meV only the states in QR are occupied.

\begin{figure*}[htb]
  \includegraphics[width=0.9\textwidth]{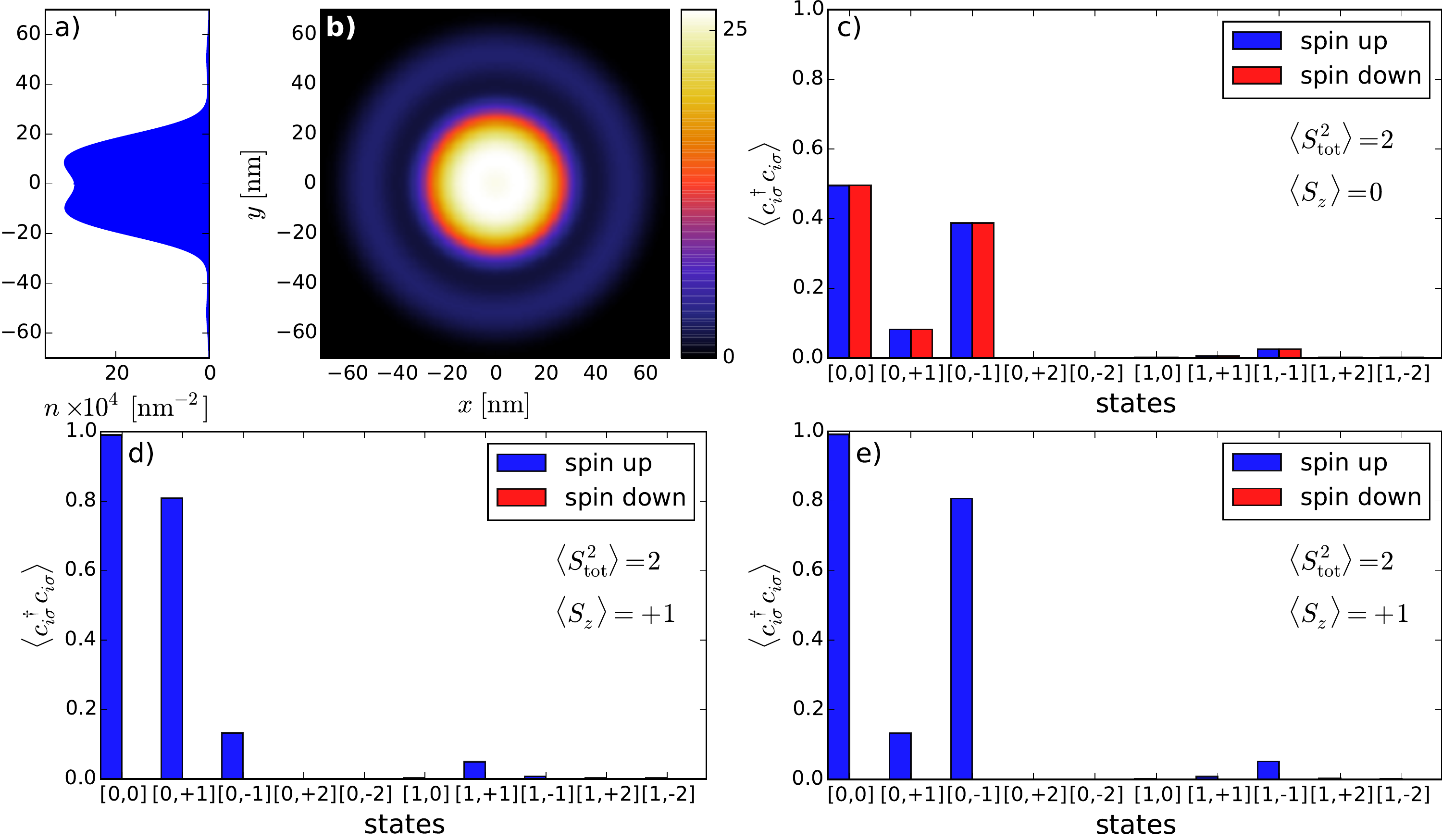}
\caption{The same as in Fig. \ref{Fig:2DRN_m4p4} for $N_e=2$, but for the first excited state with $V_\mathrm{QD}=-6$ meV. This state is six-fold degenerate and the presented electron density is averaged over all the six states. In panels c) to e) the occupancy of single-particle states is shown for half of the states in the basis. For the state with $\langle S_z\rangle=0$ [c)] there exists a counterpart with exchanged single-particle contributions $[0,+1]$ and $[0,-1]$; for the states with $\langle S_z\rangle=+1$ [d) and e)] their
  counterparts with $\langle S_z\rangle=-1$ have the same contribution.}
  \label{Fig:2DRN_m4p4_exc}
\end{figure*}

A similar evolution can also be observed for the excited states. Fig. \ref{Fig:2DRN_m4p4_exc} shows the first excited state for $V_\mathrm{QD}=-6$ meV. With increasing value of $V_\mathrm{QD}$ also the excited state is moved over to the ring part of DRN, similarly to the ground state. The evolution is presented in Supplementary Figs. \ref{Fig:Supp1} and \ref{Fig:Supp2}.

The contribution of the $2$--$3$ first single-particle functions $\varphi_i (\vec{r})$ out of $M=10$
states to $n (\vec{r})$ is usually predominant. Inclusion of e.g., $M=18$ states in \eqref{eq:field_operator} does not change practically the results. 
This last circumstance means that the interaction involves only a relatively small number of two-particle components $\ket{\varphi_i \varphi_j}$ in the resultant two-particle state $\ket{\Phi}$, at least for the lowest excited states of the system.

\subsubsection*{Three Electrons}

\begin{figure*}[htb]
  \includegraphics[width=0.9\textwidth]{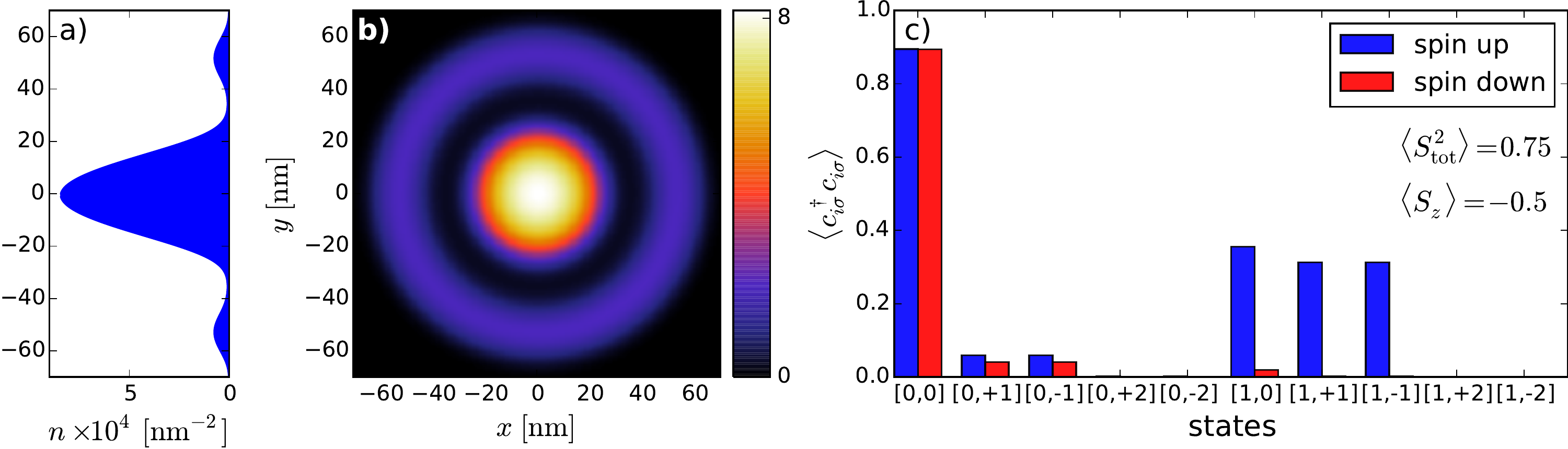}
\caption{Electronic density profile $n(x,y)$ [a), b)] and occupancy  $\bar{n}_{i\sigma} \equiv \average{\CR{i}{2} \AN{i}{2}}$ of the single-particle states included in the calculations c) for  $N_e = 3$ and $V_\mathrm{QD} = -6$ meV for the ground state. The total spin is $1/2$. Two electrons are forming a singlet and located mainly in the dot part, whereas the third (unpaired) electron is located further away, as seen by the presence of the spin-polarized subsidiary occupancy maxima in $\average{\CR{i}{2}\AN{i}{2}}$ (cf. c). The eigenenergy is $12.877$ meV. This state is degenereate, its counterpart has exchanged the occupancy of spin-up and spin-down states.}
\label{Fig:3DRN_m6_gs}
\end{figure*}

\begin{figure*}[htb]
  \includegraphics[width=0.9\textwidth]{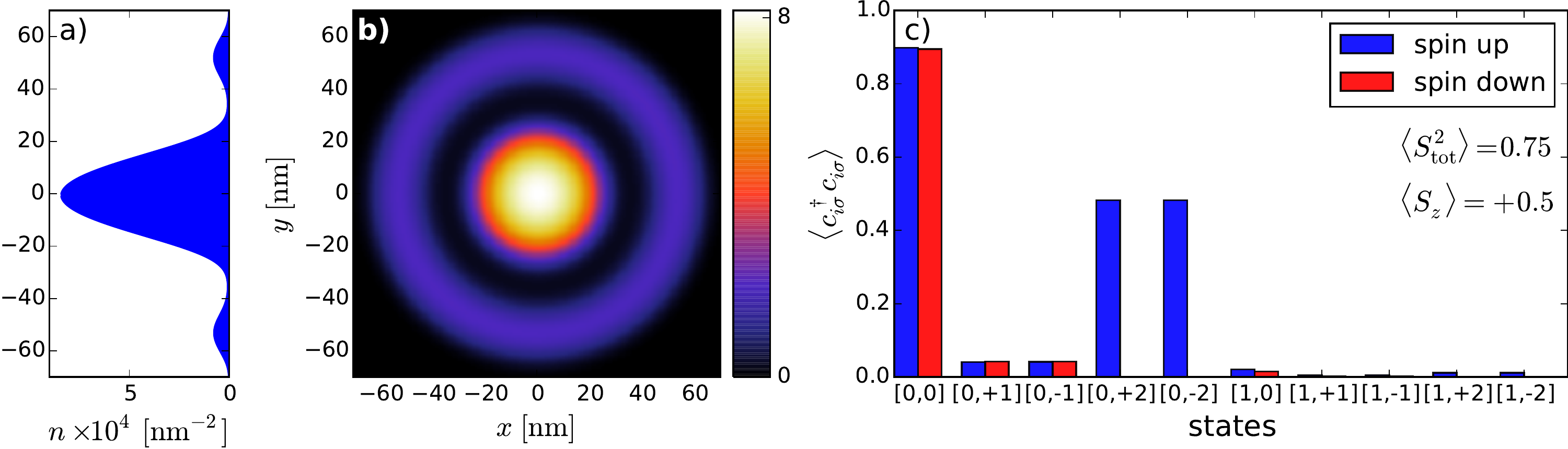}
  \caption{The same as in Fig. \ref{Fig:3DRN_m6_gs} for $N_e=3$, but for the first excited state. The electronic density is almost exactly the same, but the occupancy $\bar{n}_{i\sigma}$ of the single-particle states is different. The state eigenenergy is $13.579 \approx 1 mRy$ meV.}
  \label{Fig:3DRN_m6_ex}
\end{figure*}

Next, we present electronic density, as well as the squares of the total spin and the spin component along an arbitrarily selected $z$ axis for the ground and the first excited states of DRN
for 3 electrons (cf. \cref{Fig:3DRN_m6_gs,Fig:3DRN_m6_ex}). The ground state
is the state with the total spin $S=1/2$ ($\average{\vec{S}_{tot} ^2}=3/4$) for $V_\mathrm{QD}<3$ meV or $S=3/2$ ($\average{\vec{S}_{tot} ^2}=15/4$) for $V_\mathrm{QD} \geq 3$ meV.
For the high spin state, a redistribution of the density $n(\vec{r})$ into the products of single-particle component is more involved, as one would expect,
whereas for $S=1/2$ the state is composed of the dominant pair-singlet state and the third electron in a higher orbital with 
the dominant ring contribution.

Parenthetically, it would be interesting to calculate the transport properties via tunneling through the DRN with $N_e=2$ as this would involve cumbersome intermediate state
with $N_e=3$. Depending on $V_\mathrm{QD}$, the tunneling probability is allowed (for $S=0$) and substantially suppressed when $S=1$ (in applied field).
Such effects should be analyzed separately as they involve an analysis of electronic transitions between the many-electron states.

\subsection*{Coulomb-interaction parameters}
\begin{figure}[h!]
 \includegraphics[width=.65\linewidth]{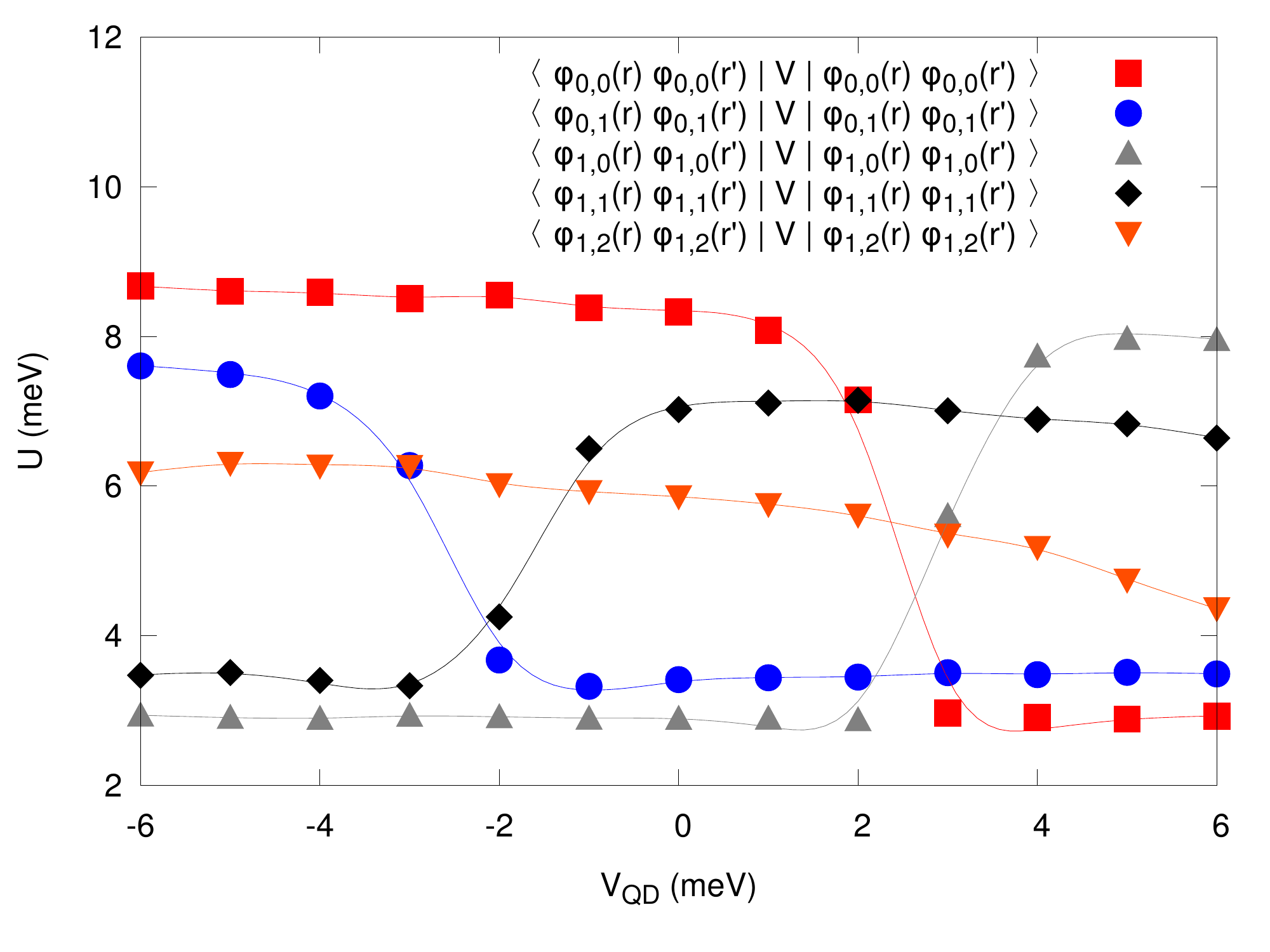} 
 \caption{Values of the Hubbard (intrasite) repulsion $U_{i} \equiv V_{iiii}$ vs the tuning parameter  $V_\mathrm{QD}$  for different states, as marked. These values are in some cases comparable
 to the single-particle energy, so the interelectronic correlations are very important then. The continuous lines are guide to the eye to visualize the tendency of the calculated points. The nonmonotonic behavior is due
 to the level crossing depicted in Fig.~\ref{Fig:sp_energy}. }
  \label{Fig:U}
\end{figure}

\begin{figure}[h!]
 \includegraphics[width=.45\linewidth]{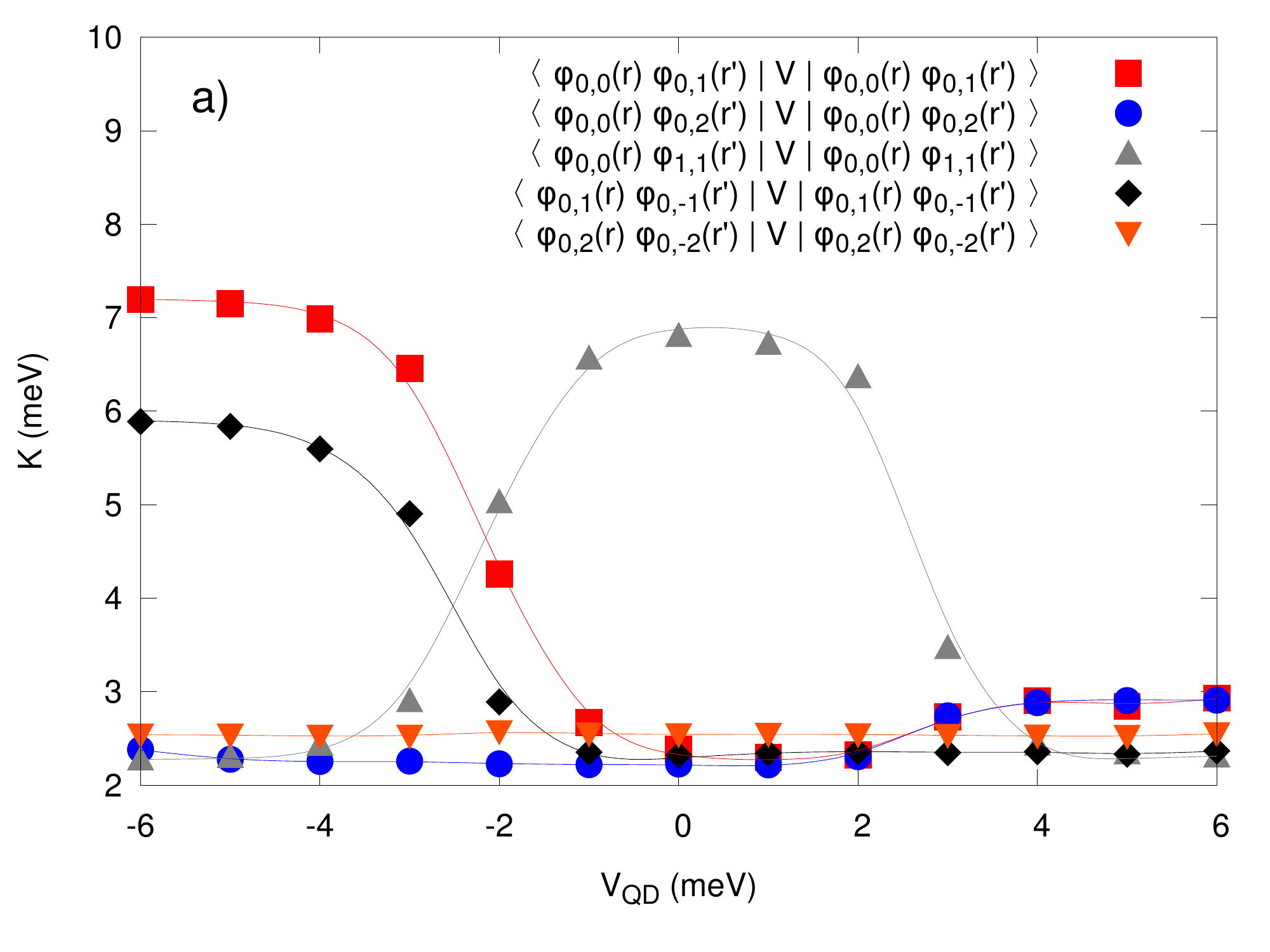} 
 \includegraphics[width=.45\linewidth]{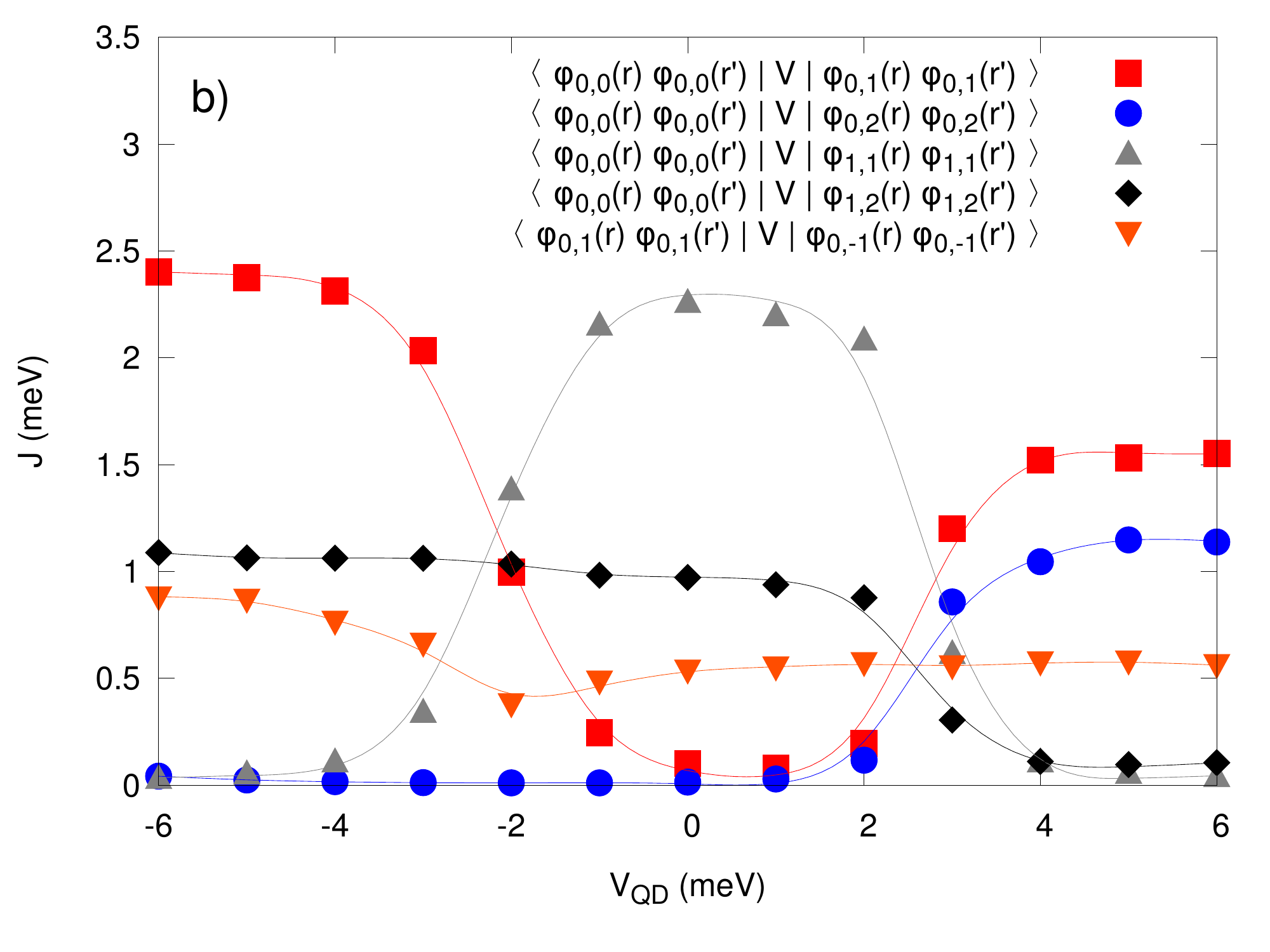}\\
 \includegraphics[width=.45\linewidth]{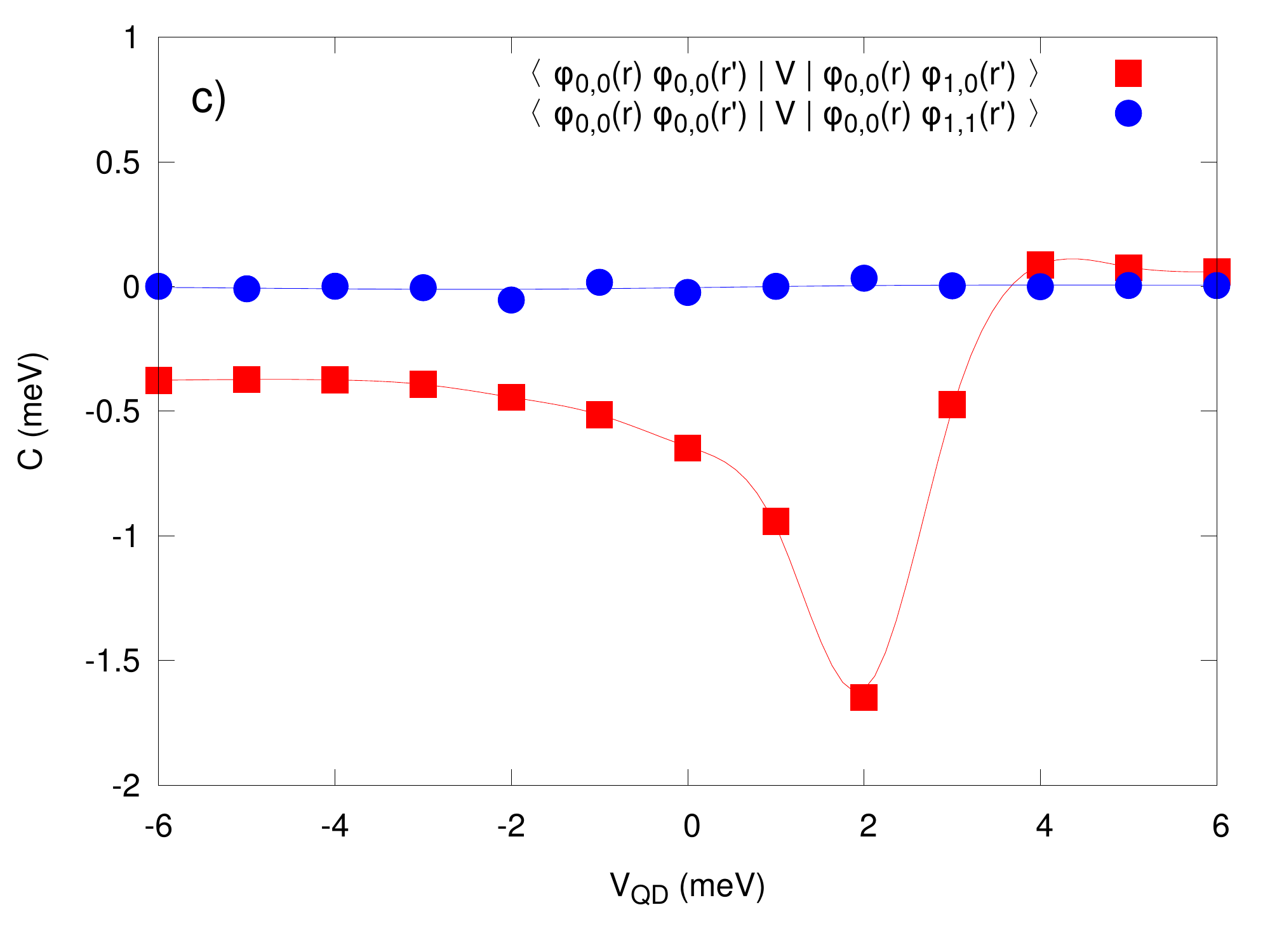} 
 \includegraphics[width=.45\linewidth]{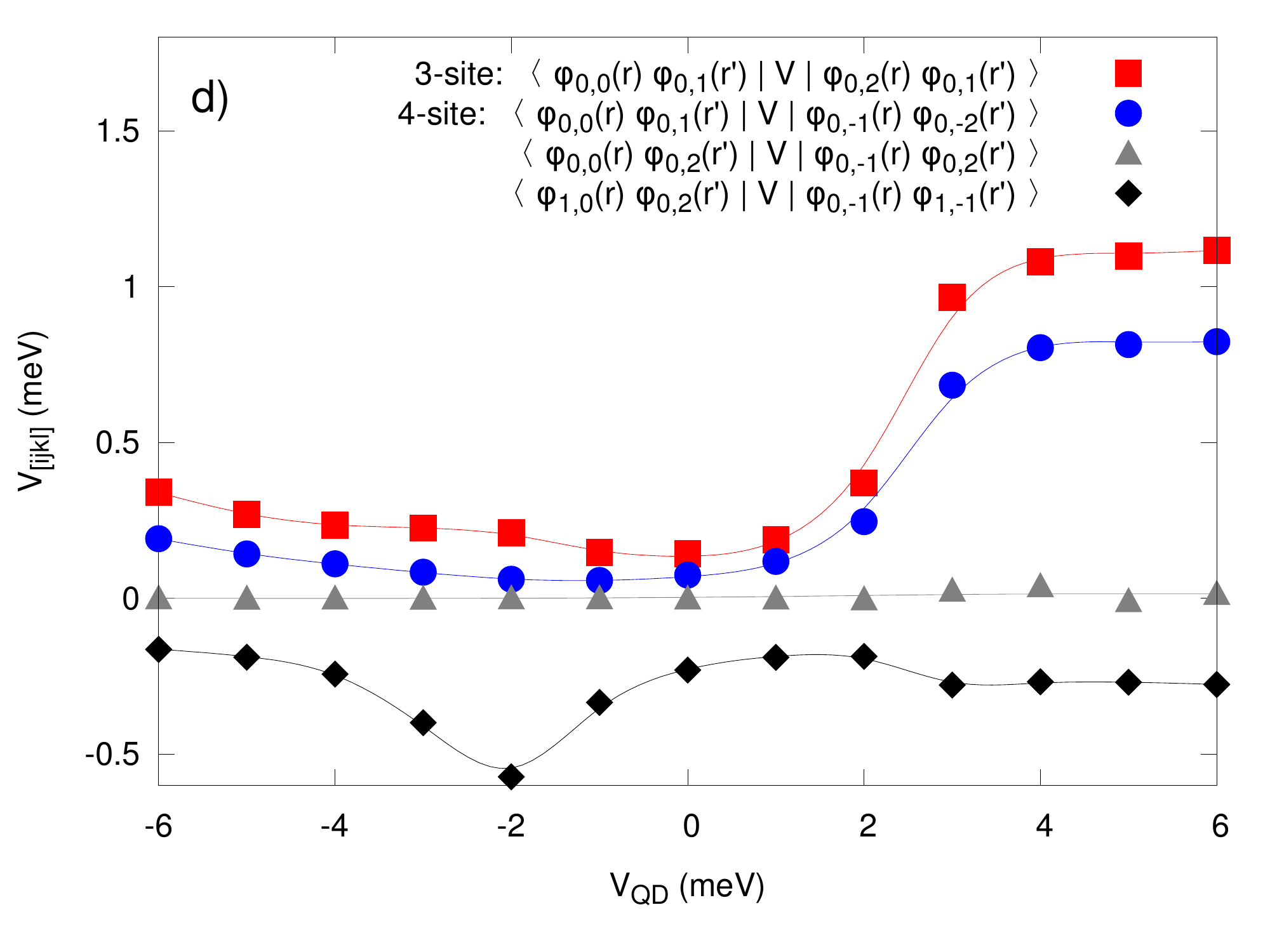} 
 \caption{\textbf{a)} Two-state (interstate) Coulomb repulsion amplitudes $K_{ij} \equiv V_{ijij}$ vs $V_\mathrm{QD}$. A rapid change of their values can be related to the left level (anti)crossing
 specified in Fig.~\ref{Fig:sp_energy} and the concomitant change of the single-particle wave-function symmetry of the corresponding states. Note also that for $V<-2$ meV
 some of the values of $K_{ij}$ can even become comparable to those of $U_i$.
 \textbf{b)} Exchange integral $J_{ij} \equiv V_{ijji}$ vs QD potential. The convention is that $J_{ij}>0$ denotes the ferromagnetic spin-spin exchange interaction.
 \textbf{c)} Selected correlated hopping amplitudes $C_{ij} \equiv V_{ijjj}$ vs QD potential. The terms containing this parameter (the sixth term in \eqref{eq:hamiltonian_special})
 lead to the nonorthogonality of the starting single-particle basis $\{ \varphi_i (\vec{r}) \}$, i.e., to the hopping between those states with a double occupancy
 in either initial of final states.
 \textbf{d)} Selected three- and four-site $V_{[ijkl]}$ parameters vs QD potential. They become comparable to $J_{ij}$ for $V_\mathrm{QD}>2$ meV, where the second level
 crossing appears, particularly when one takes into account the circumstance that the number of, e.g., four-state terms is relatively large.
 The continuous lines are guide to the eye.
 }
  \label{Fig:KJC}
\end{figure}

We now turn to the most basic aspect of our present work. Namely, we calculate all possible microscopic interaction parameters
$V_{ijkl}$ appearing in \eqref{eq:hamiltonian}. Those parameters appearing in the microscopic parameters reflect various quantum
processes encoded in the starting Coulomb repulsion. This procedure should allow us to determine a coherent
and exact many-particle physical picture with concomitant information concerning the importance of various classes of interaction
terms, as expressed via the respective one-, two-, three-, and four-state terms. We start by rewriting the starting Hamiltonian
\eqref{eq:hamiltonian} to the following form
\begin{align}
 \label{eq:hamiltonian_special}
 \hat{\mathcal{H}} &= \sum_{i,\sigma}\epsilon_{i} \NUM{i\sigma}{0} +\sum_{i}U_{i}\NUM{i}{1} \NUM{i}{-1}\\\notag
&- \sum_{i \neq j} J_{ij} \vec{S}_i \cdotp \vec{S}_j + \frac{1}{2}\sum_{i\neq j}\left(K_{ij} - \frac{1}{2}J_{ij}\right)\NUM{i}{0} \NUM{j}{0}\\\notag
& + \sum_{i \neq j}J_{ij} \CR{i}{1} \CR{i}{-1} \AN{j}{-1} \AN{j}{1} + \sum_{\sigma,i\neq j}C_{ij} \NUM{i}{2} \left( \CR{i}{-2}\AN{j}{-2} + \CR{j}{-2}\AN{i}{-2} \right) \\\notag
& + \frac{1}{2} \sum_{[i j k l], \sigma, \sigma'} V_{ijkl} \CR{i}{2} \CR{j \sigma'}{0} \AN{l \sigma'}{0} \AN{k}{2}, 
\end{align}
where the first 6 terms represent one- and two-state interactions\cite{Spalek1,Spalek2}, respectively, and 
$\sum_{[ijkl]}$ refers to sum over indices with at least three of them being different. The first question relates to the
magnitude of the intrasite Hubbard interaction, $U_i \equiv V_{iiii}$ (cf. Fig.~\ref{Fig:U} and Table~\ref{tab:U} in Supplem. Material), the generic term in the Hubbard model, as compared
to the inter-state repulsion $K_{ij} \equiv V_{ijij}$  (cf. Fig.~\ref{Fig:KJC} and Table~\ref{tab:K} in Suppl. Mat.), the exchange energy 
$J_{ij} \equiv V_{ijji}$ (cf. Fig.~\ref{Fig:KJC} and Table~\ref{tab:J}), and the so-called correlated hopping $C_{ij} \equiv V_{ijjj}$  (cf. Fig.~\ref{Fig:KJC}
and Table~\ref{tab:C}). In the present situation, the inclusion
of three- and four-index interaction parameters $V_{[i j k l]}$  (cf. Fig.~\ref{Fig:KJC} and Table~\ref{tab:V}) 
 is of the crucial importance, as these parameters are usually omitted in the models describing various quantum devices. The reason for including them is due to the circumstance that in a few-electron system there is no screening and thus, in principle, all the terms may become relevant. In any case, on the example of DRN we can see explicitly the role of \textbf{all} consecutive terms, what is, in principle, of fundamental importance for a reliable modeling of the nanodevices.
 These last terms proved to be nonnegligible as shown in \cref{Fig:dens_diff,Fig:dens_diff_VQD} (cf. also Table~\ref{tab:V}),
 and can become even of comparable magnitude to the exchange energy.

 Visible in most of the cases in \cref{Fig:U,Fig:KJC} are the rapid changes of the microscopic parameters
 which coincide with the single-particle level-crossings observed in the single-particle levels (cf. Fig.~\ref{Fig:sp_energy}), but these do not influence in any essential manner
 the resultant many-particle picture, as may be explicitly seen in Fig.~\ref{Fig:sp_energy}, where we observe
 a smooth evolution with changing $V_\mathrm{QD}$.


\subsection*{Two-state versus the three and four-state interaction contributions}

\begin{figure}[h!]
 \includegraphics[width=.45\linewidth]{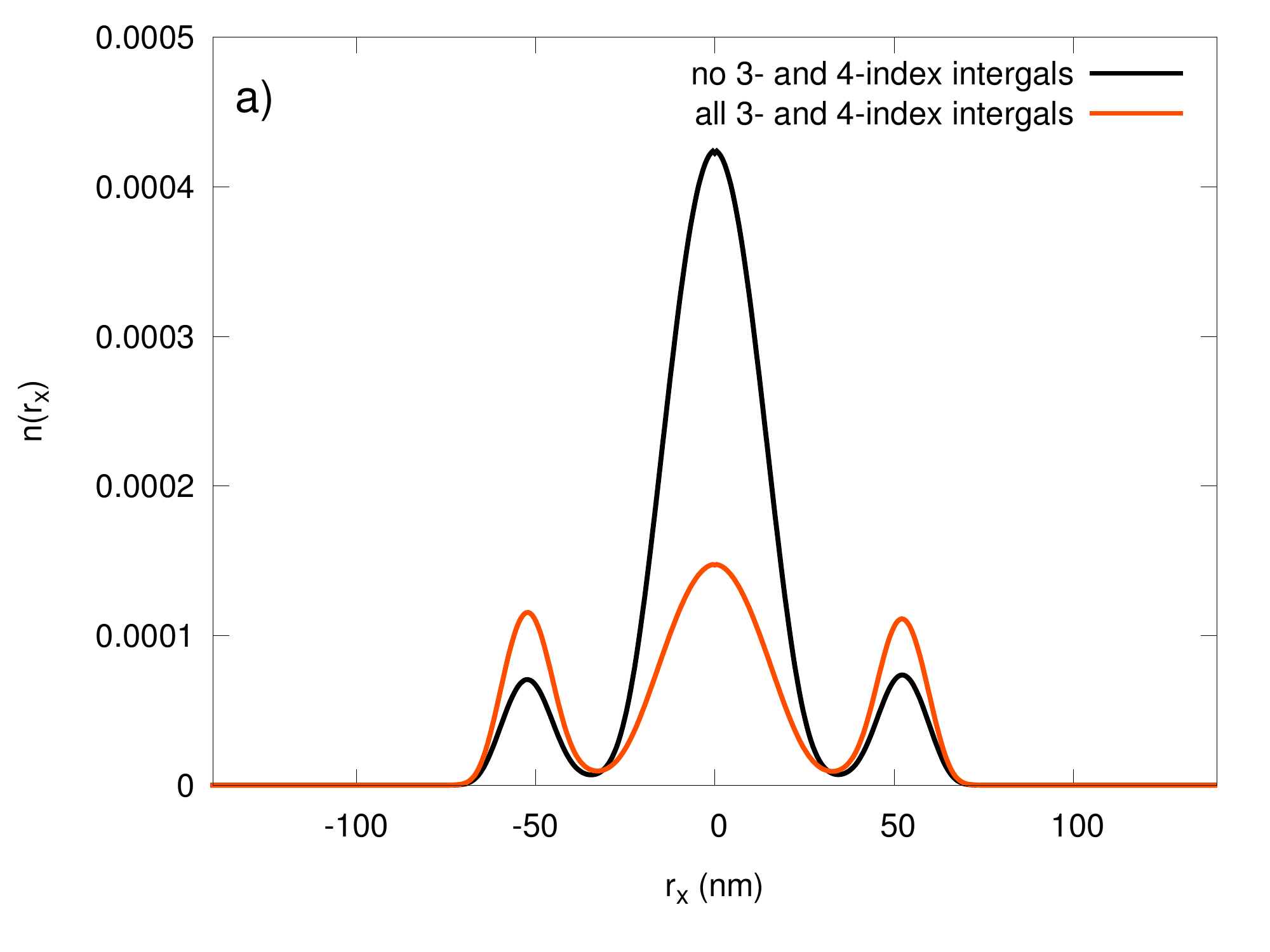}
 \includegraphics[width=.45\linewidth]{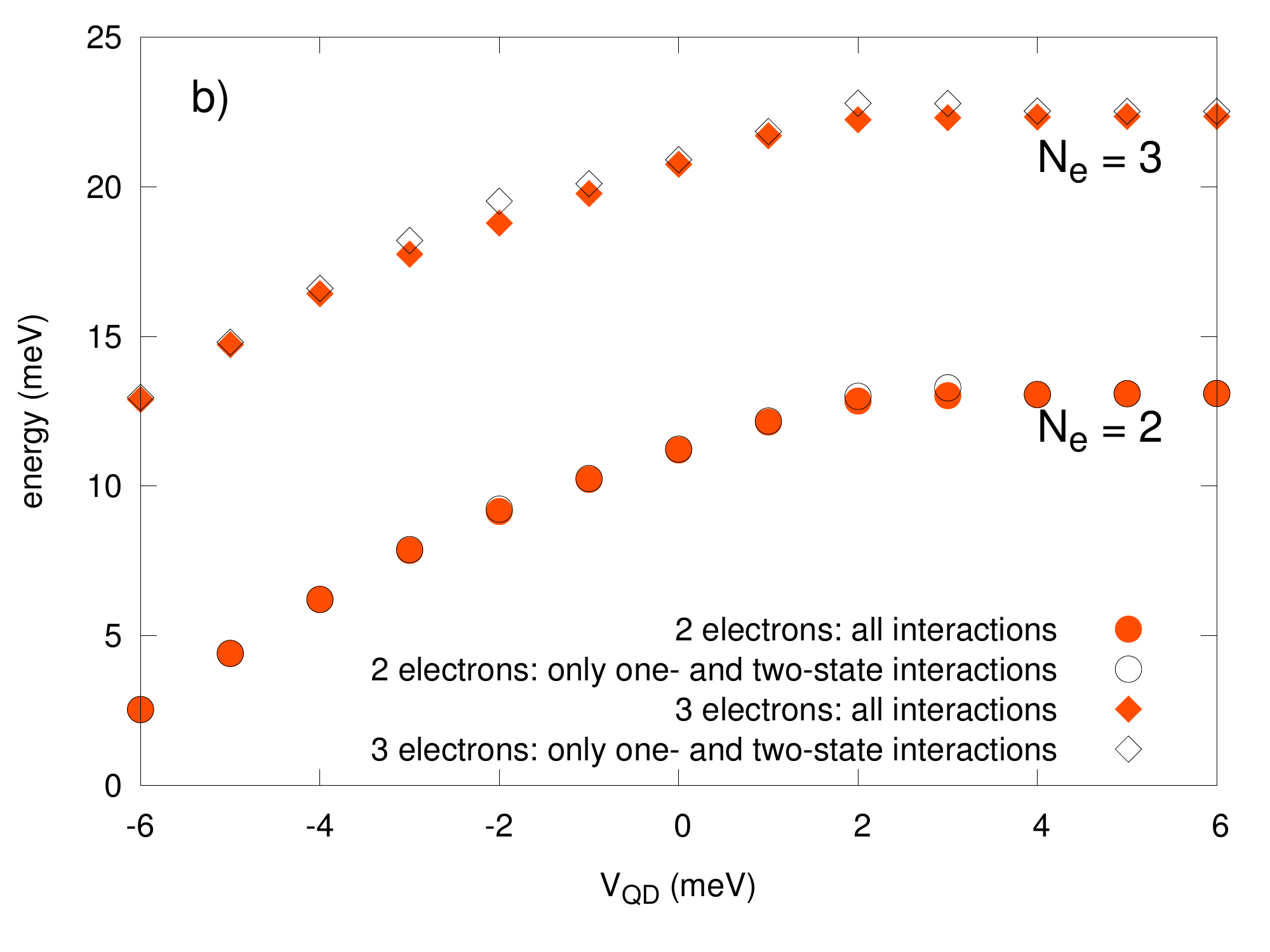} 
 \caption{\textbf{a)} Electronic density along the $x$ axis (for $y\equiv 0$) for the case of $N_e=2$ and $V_\mathrm{QD}=2$ meV, with no 3- and 4-indices integrals included (black), as compared to that coming from our present approach 
 which includes all the microscopic parameters (orange). In the former case, the density is reduced essentially to the dot region. \textbf{b)} The ground state energies
 for $N_e=2$, and $3$ without and with the 3- and 4-state interactions included. The role of the latter is of secondary importance for these quantities.}
  \label{Fig:dens_diff}
\end{figure}

We illustrate next the role of the pairwise vs. 3- and 4-state interactions with their paramters displayed in Figs. \ref{Fig:U} and \ref{Fig:KJC}. For that purpose, we draw in Fig.~\ref{Fig:dens_diff}a 
the exemplary profile of the electron density cross section $n(r_x)$ for $r_y\equiv 0$,for $N_e=2$ without and with the 3- and 4-state interactions included. The role of the latter terms is essential.
As expected, with those interactions included, the electrons are pushed to the ring region in that situation. On the contrary, the role of the 3- and 4-state
terms is not so crucial when evaluating the ground state energy (cf. Fig.~\ref{Fig:dens_diff}b). Therefore, one sees that the 3- and 4-state interactions
will be of primary importance when evaluating the matrix elements between the states.

\begin{figure}[h!]
 \includegraphics[width=.45\linewidth]{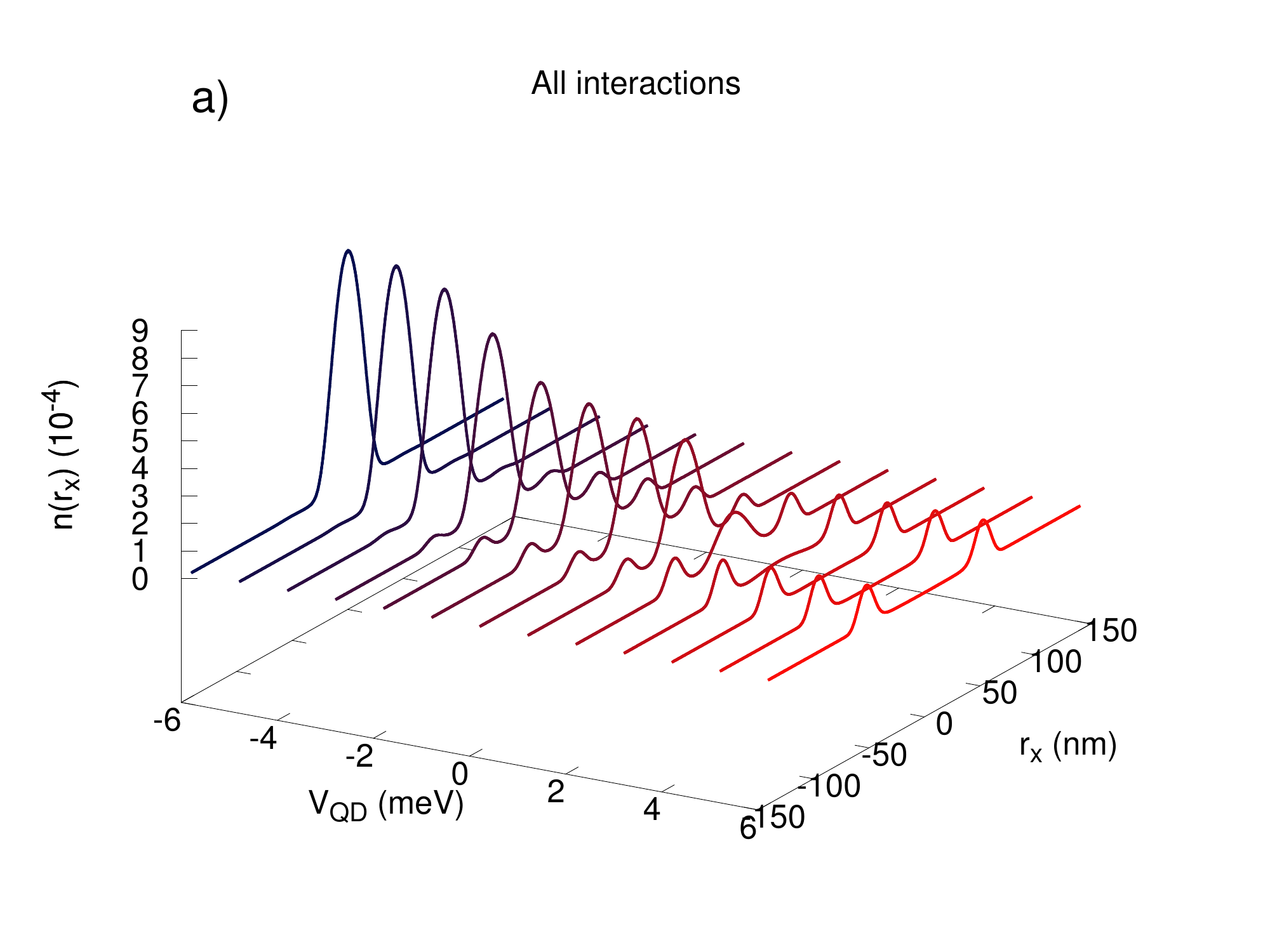}
 \includegraphics[width=.45\linewidth]{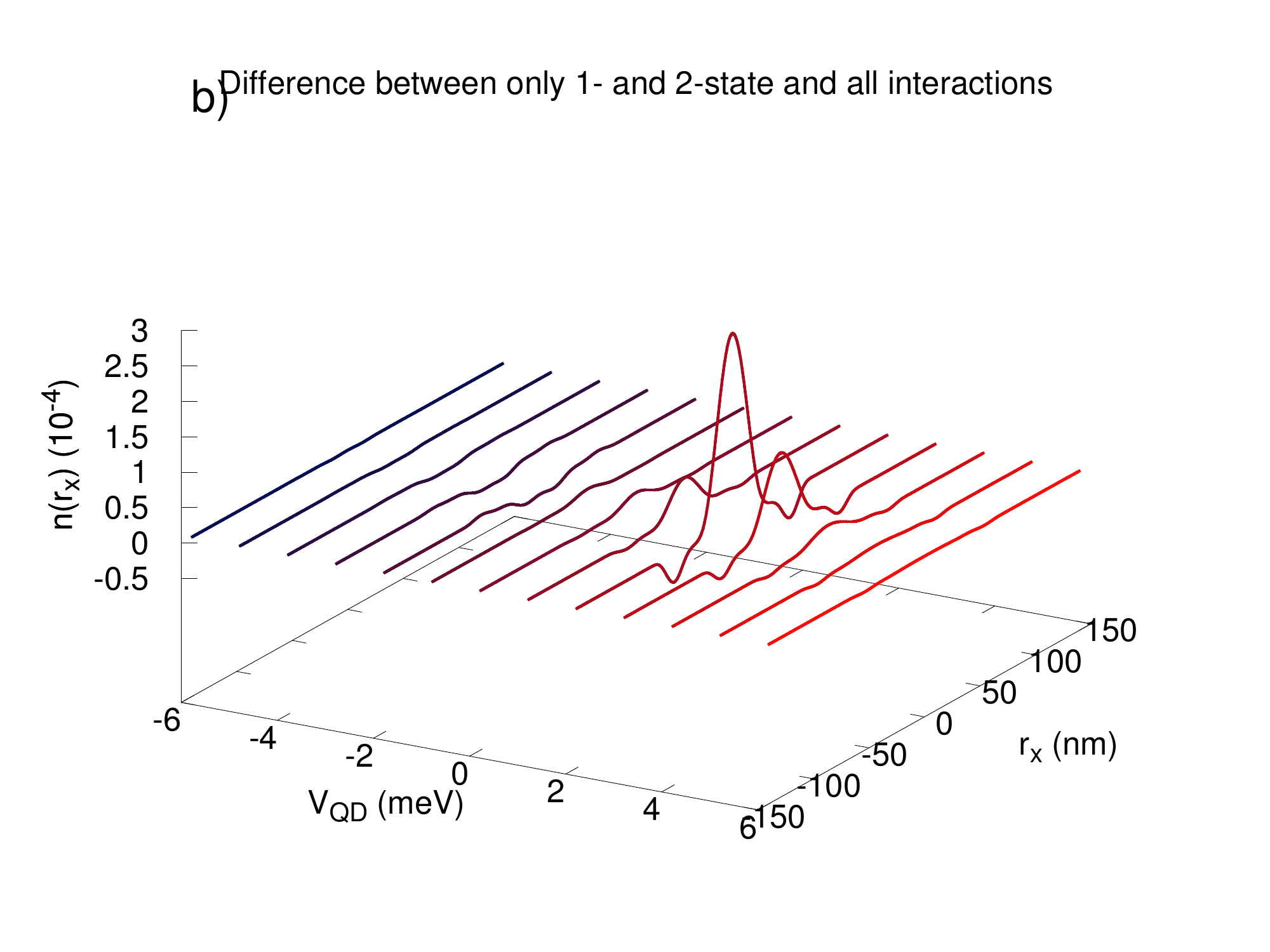} 
 
 \caption{\textbf{a)} Electronic density along the $x$-axis, i.e., on the $V_\mathrm{QD}-r_x$ plane ($r_y \equiv 0$), with all the interactions included. \textbf{b)} Difference in the density between the situation
 with the pairwise interactions taken only and that with all the interaction terms taken into account. The difference is the largest in the $QD$ region and for $V_\mathrm{QD}$ at and around zero.}
  \label{Fig:dens_diff_VQD}
\end{figure}

To determine explicitly the role of the three- and four-state interaction terms we have plotted in Fig.~\ref{Fig:dens_diff_VQD} the particle density
profile with and without inclusion of them. We see that their role is crucial. Note that each of the curves has the same area equal to $2$ (the number $N_e$).
The apparent inequivalence arises from the circumstance that the ring part encompasses effectively a larger volume (here only 
a single cross-section $n(r_x)$ is plotted). So all the interaction terms contribute in a nontrivial manner to the many-particle wave-function
engineering! Also, one can compose a resultant many particle state out of the products of the single-particle basis states and the leading terms are
\begin{align}
 \label{eq:wf_full}
 \ket{\Psi_0} \equiv \sum_{ij\sigma \sigma'} A_{ij\sigma\sigma'} \CR{i}{2} \CR{j}{3} \ket{0} \approx \big( -0.6704 ( \CR{0,1}{1}\CR{0,0}{-1} + \CR{0,0}{1}\CR{0,1}{-1} ) - 0.2890 \CR{0,0}{1}\CR{0,0}{-1} + \cdots \big) \ket{0}.
 \end{align}
The complete list of the leading coefficients $A_{ij\sigma\sigma'}$ for the ground state spin singlet is provided in Table~\ref{tab:wf_coeffs}. Note that their values are the same for the components $\CR{i}{2} \CR{j}{3}$
and $\CR{j}{2} \CR{i}{3}$, of that singlet state. Essentially, the decomposition \eqref{eq:wf_full} with the complete list of the coefficients (cf. Table~\ref{tab:wf_coeffs}) provides the same type of expansion as that appearing in the Configuration Interaction (CI) method \cite{SzaboOstlund}. Here, a particular combination of the pair products of the creation operators represents a single Slater determinant of the single-particle wave functions and the respective numerical values of the coefficients describe the weight of each two-particle Slater determinant state. From Table~\ref{tab:wf_coeffs} we see that only limited number of such states matter in this (and other) cases. This means that if the number $M$ of single-particle states in \eqref{eq:field_operator} is selected properly, the obtained results for many-particle states and their eigenvalues can be achieved to a very high accuracy. Here, it has been sufficient to choose $M=10$ for $N_e=2, \ 3$. For the state \eqref{eq:wf_full} this
results in having 24 leading coefficients listed in Table~\ref{tab:wf_coeffs}, i.e., the state can be represented
well by 24 component states composing that state.
For larger values of $N_e$, the method is also workable, but the value of $M$ must be selected with care.
\begin{table*}[t!]
\caption{Values of the leading coefficients $A_{ij\uparrow\downarrow}$ ($>10^{-4}$) for the case $V_\mathrm{QD}=0$ meV, and for the two-particle state \eqref{eq:wf_full} for different pairs $(i,j)$ of states composing
this state. }
\label{tab:wf_coeffs}
\resizebox{\textwidth}{!}{
 \begin{tabular}{c||r|r|r|r|r|r|r|r|r|r|}
 \mc{1}{c}{} & \mc{1}{c}{$n=0,\ l=0$} & \mc{1}{c}{$n=0,\ l=1$} & \mc{1}{c}{$n=0,\ l=-1$} & \mc{1}{c}{$n=0,\ l=2$} & \mc{1}{c}{$n=0,\ l=-2$} & \mc{1}{c}{$n=1,\ l=0$} & \mc{1}{c}{$n=1,\ l=1$} & \mc{1}{c}{$n=1,\ l=-1$} & \mc{1}{c}{$n=1,\ l=2$} & \mc{1}{c}{$n=1,\ l=-2$} \\\hline
$n=0,\ l=0$ & -0.2890 & & -0.0005 & & & -0.6704 & & -0.0002 & & \\\hline 
$n=0,\ l=1$ & & 0.0652 & & & & & -0.0450 & & & \\\hline 
$n=0,\ l=-1$ & -0.0005 & & 0.0653 & & & & & -0.0451 & & \\\hline 
$n=0,\ l=2$ & & & & 0.0071 & & & & & -0.0049 & \\\hline 
$n=0,\ l=-1$ & & & & & 0.0071 & & & & & -0.0049 \\\hline 
$n=1,\ l=0$ & -0.6704 & & & & & -0.0149 & & & & \\\hline 
$n=1,\ l=1$ & & -0.0450 & & & & & 0.0169 & & & \\\hline 
$n=1,\ l=-1$ & -0.0002 & & -0.0451 & & & & & 0.0169 & & \\\hline 
$n=1,\ l=2$ & & & & -0.0049 & & & & & 0.0044 & \\\hline 
$n=1,\ l=-2$ & & & & & -0.0049 & & & & & 0.0045 \\\hline 
\end{tabular}
}
\end{table*}

\subsection*{Electronic transition from the ground to the excited state}
\begin{figure*}[h!]
\includegraphics[width=0.9\textwidth]{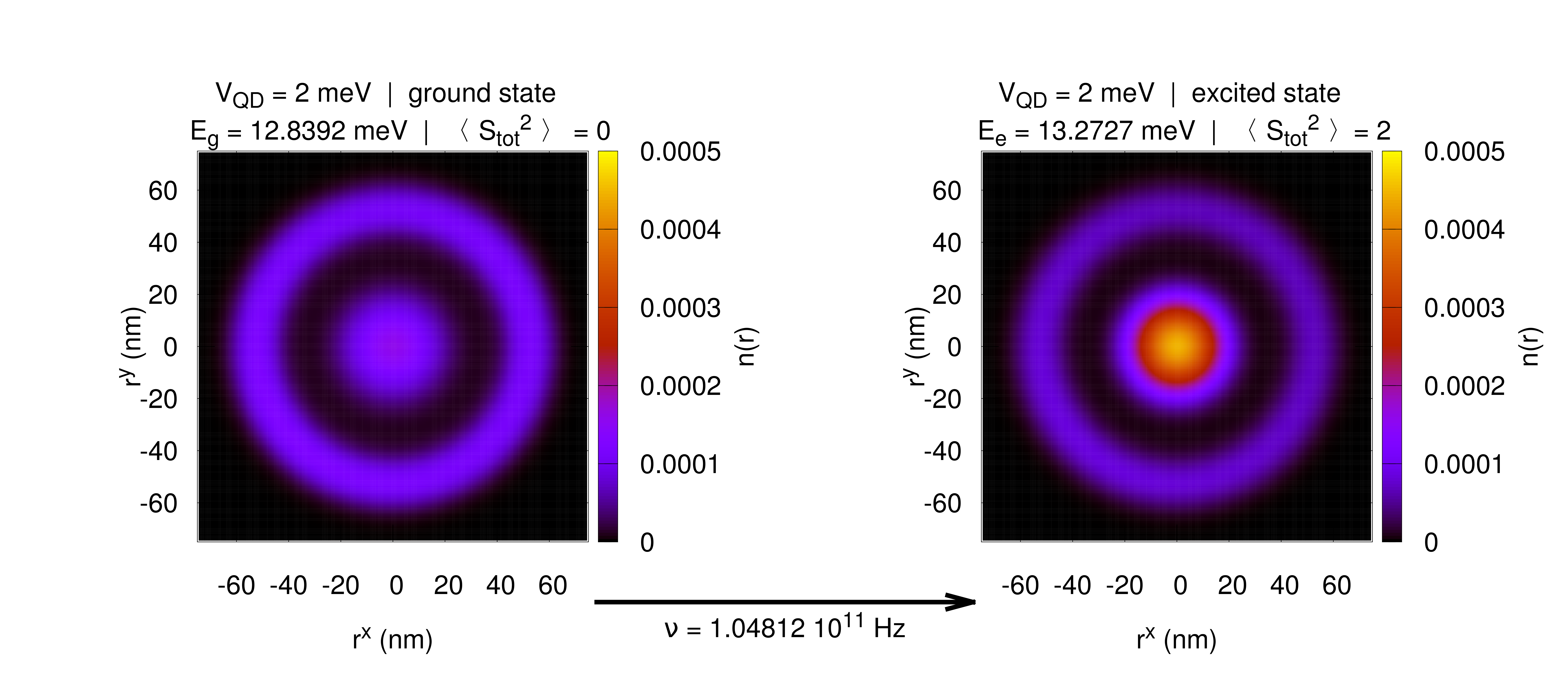}
\caption{Change in the overall electronic density for two electrons in DRN, for $V_\mathrm{QD}=2$ meV, after absorption of a photon of frequency $\nu = 1.05 \ 10^{11} \ Hz$.
Note that this excitation is allowed as the change of respective angular orbital and spin momenta are $\Delta L_{tot} = 0$ and  $\Delta S_{tot} = 1$.}
\label{Fig:excitation}
\end{figure*}

To flash on the importance of the system behavior, we examine the possibility of changing the state of electrons in DRN 
via an intraband photo-excitation for $N_e=2$. From the experimental point of view, the possibility of changing the probability of electrons to be in QD or QR is of importance.
This can be realized by a microwave radiation absorption, as illustrated in Fig.~\ref{Fig:excitation}. The selection rules are fulfilled
as we are starting from the state $L_{tot}=S_{tot}=0$ and ending in the state $L_{tot}=0$, $S_{tot}=1$, where $L_{tot}$ and $S_{tot}$
represent the orbital and spin state of the system, respectively.

A detailed analysis of the interstate transition drawn in Fig.\ref{Fig:excitation} may important principal information about 3- and 4-state interactions. Namely, by studying
DRN systems of a variable size, one should see their diminishing role with the increasing system size. Such measurements when performed, can be readily analyzed
within the exact solution provided here (the codes for the analysis of DRN for $N_e>1$ are available at \url{https://bitbucket.org/azja/qmt}).

\section*{Outlook}
\label{sec:outlook}

In this paper we have addressed in a rigorous manner the question of importance of the interelectronic interactions/correlations in nanodevices
(on example of DRN). The cases tackled explicitly were those with $N_e=2$ and $N_e=3$ electrons. We have calculated all relevant
interaction parameters and their evolution with the tuning parameter, which in this case is the relative potential $V_\mathrm{QD}$
of the quantum dot (QD) with respect to that for the ring (taken as zero). We have proved explicitly that practically \textbf{all}
relevant interaction terms are important, as they change essentially the shape of the multiparticle wave function.
The situation may depend on the size of DRN system, i.e., it may gradually become not so important with the increasing DRN size.
Such feature could be tested experimentally

To test further the role of many-particle interactions, one can follow the two principal directions. First, the determination
of the states in an applied magnetic field and in this manner see the evolution/crossing of many-particle states as the field increases. This topic
can become quite interesting as the transition between low and high spin states may turn out then to be quite nontrivial.
Second, the charge transport/tunneling processes through DRN can be nontrivial as they should also be connected with the total
spin values change when applied field/$V_\mathrm{QD}$ are altered. It has already been demonstrated that in the single electron regime
the system can be applied as a switching device (transistor) \cite{Kurpas1}. Taking into account the possibility of controlling many-particle 
states, such situation would allow for manipulating the spin-dependent coupling between the DRN and the leads. This, in turn, opens a new area of applications, 
also in single spintronics, e.g., as spin valves or spin filters. We should see a progress along theses lines soon.

Finally, as mentioned above, one could also vary the system size and see the evolution of the relative roles of single-particle vs. many-particle
contributions to the total energy. The latter part will gradually become less important with the increasing system size.
In this manner, the DRN system may be useful for not only single-electron, but also for many-particle wave-function engineering and associated with it total-spin value changes.

\section*{Acknowledgment}
\label{sec:Acknowledgment}
Three authors (AB, APK, and JS) were supported by the National Science Centre (NCN) through Grant MAESTRO, No. DEC-2012/04/A/ST3/00342,
whereas the others (AG-G, EZ, and MMM) by Grant No. DEC-2013/11/B/ST3/00824. The authors are also grateful to Dr. Paweł Wójcik and his student Szymon Olejak
for sharing with us their unpublished results.

\bibliography{biblio}

\section*{Contributions}
\label{sec:contributions}

M. M. M. and J. S. posed the problem and method of approach. A. G.-G. calculated the single-particle wave functions.
A. B. and A. P. K. contributed equally to the numerical calculations of the multi-particle states. J. S. prepared
the first and the final versions of the paper. A. B., A. P. K., A. G.-G., E. Z., M. M. M., and J. S. contributed to its final form.

\section*{Financial interests}
\label{sec:finantialinterests}

The authors declare no competing financial interests.
\label{PAG:end}


\newpage
\beginsupplement

\section*{Supplementary Information}
{\LARGE Dot-ring nanostructure: Rigorous analysis of many-electron effects} \\ \\
Andrzej Biborski$^1$, Andrzej P. K\k{a}dzielawa$^2$, Anna Gorczyca-Goraj$^3$, El\.zbieta Zipper$^3$, Maciej M. Ma\'{s}ka$^3$, and
J\'{o}zef Spa\l{}ek$^{1,2}$
\begin{itemize}
    \item [$^1$] Akademickie Centrum Materiałów i Nanotechnologii,
AGH Akademia Górniczo-Hutnicza,
Al. Mickiewicza 30, Krak\'{o}w, Poland
\item[$^2$] Instytut Fizyki im. Mariana Smoluchowskiego, Uniwersytet Jagielloński, ul. Łojasiewicza 11, PL-30-348 Krak\'{o}w, Poland
\item[$^3$] Instytut Fizyki, Uniwersytet Śl\k{a}ski, ul. Uniwersytecka 4, PL-40-007 Katowice, Poland
\end{itemize}

\appendix

\section{Starting basis of single-particle wave-functions}

The shapes of the ten selected single-particle real wave functions $\varphi_{nl} (\vec{r})$ forming a trial 
basis for the definition of the field operators in \eqref{eq:field_operator}
are characterized below in Figs.~\ref{Fig:R0}~and~\ref{Fig:R1}. The method of calculating those 
wave-functions relies on solving the wave equation for a single electron
for the potential energy depicted in Fig.~\ref{Fig:drn_str}. The details and the method accuracy 
is discussed elsewhere \cite{Zipper1,Kurpas1,Kurpas2014}. With the rotational symmetry of the potential in the DRN plane, 
those wave functions exhibit similarity to the hydrogenic-like functions for values of $l=0, \ \pm1, \ \pm2$ respectively. 
This minimal basis with $M=10$ components in Eq. \eqref{eq:field_operator}
is sufficient to describe accurately the multiparticle states for $N_e=2$ and $3$ electrons analyzed in main text. A subsequent enlargement of the starting basis to $M=18$ functions did not influence the accuracy of the presented results.

\begin{figure}[p]
 \includegraphics[width=.44\linewidth]{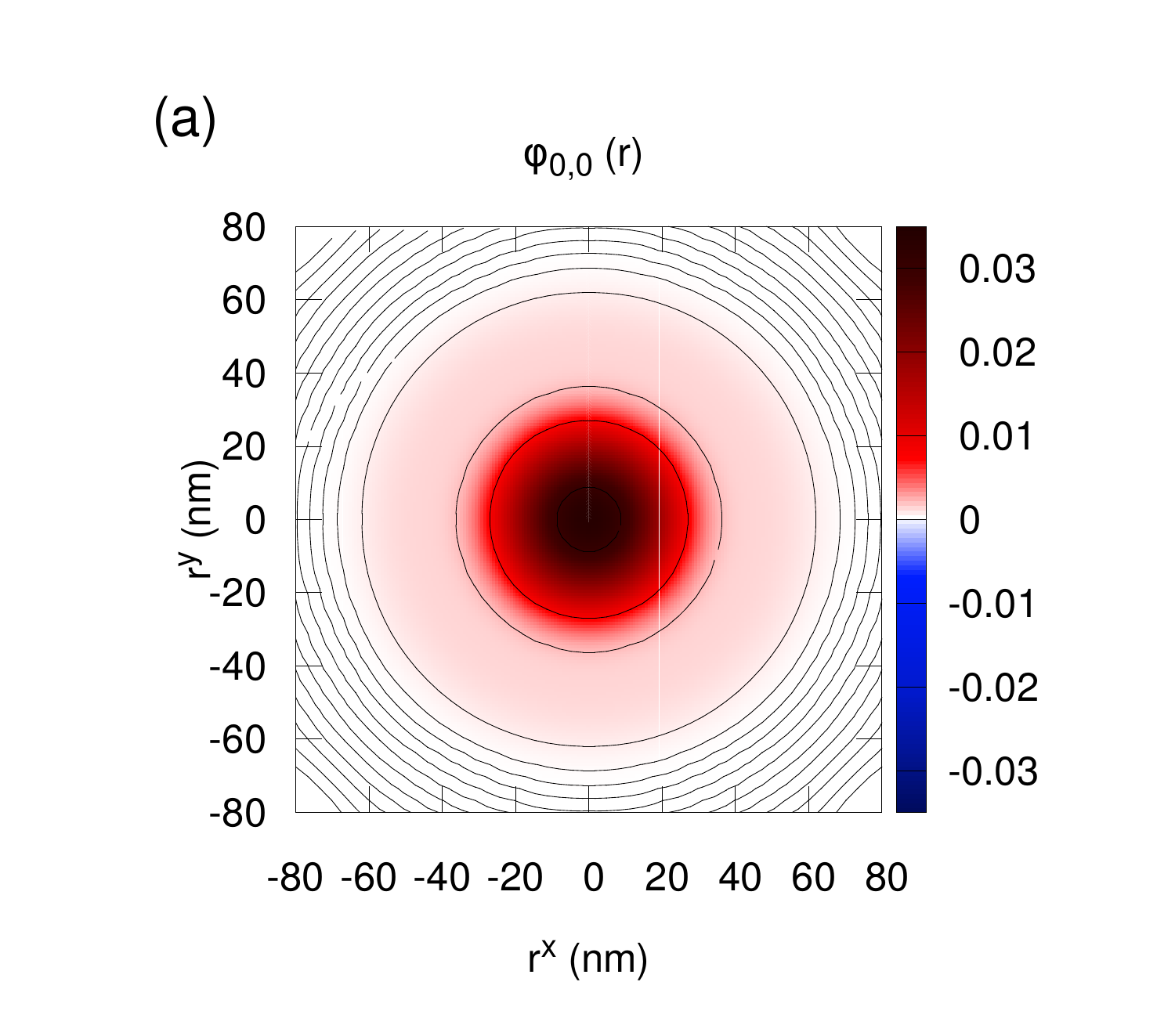} \\
 \includegraphics[width=.44\linewidth]{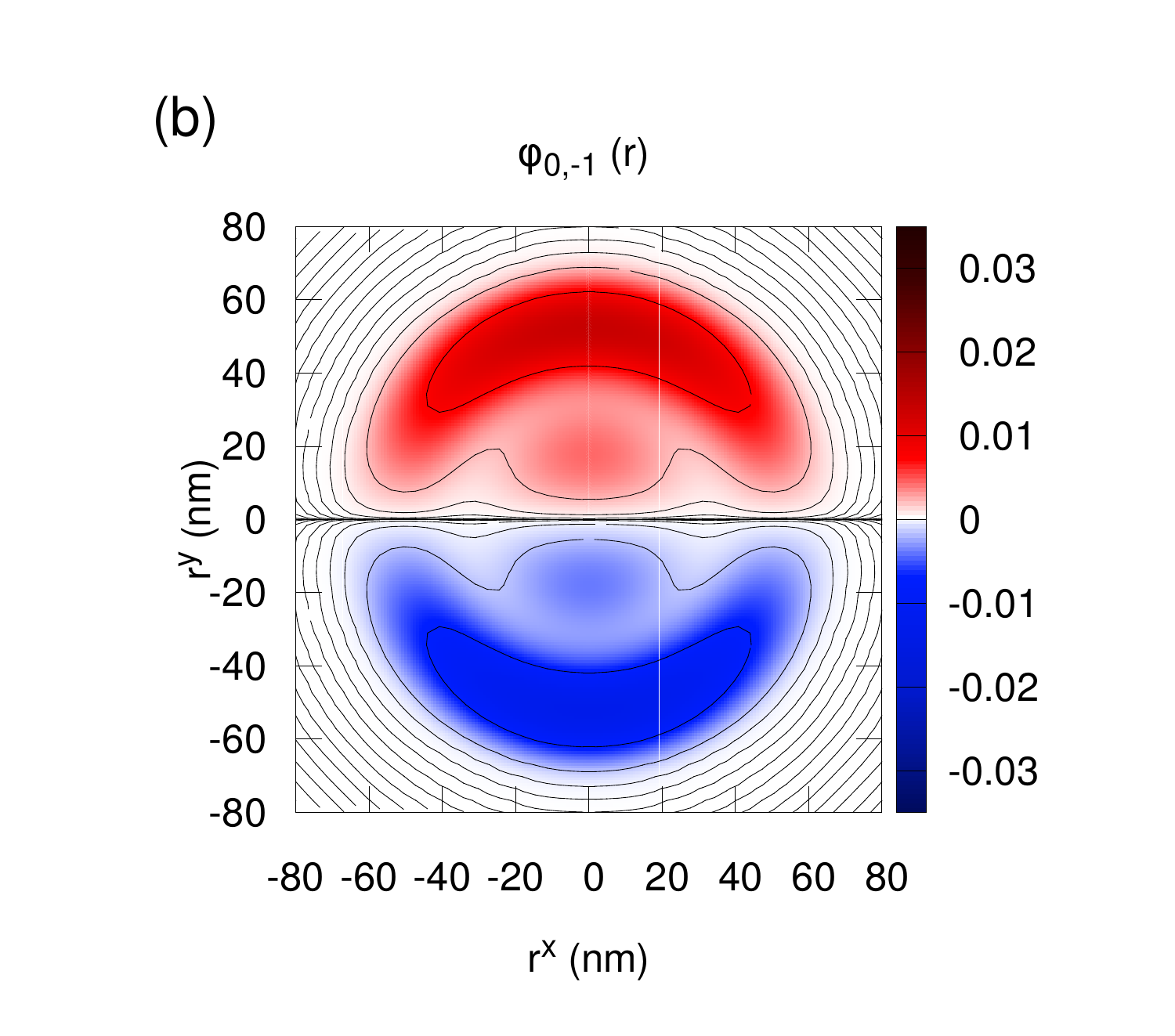}
 \includegraphics[width=.44\linewidth]{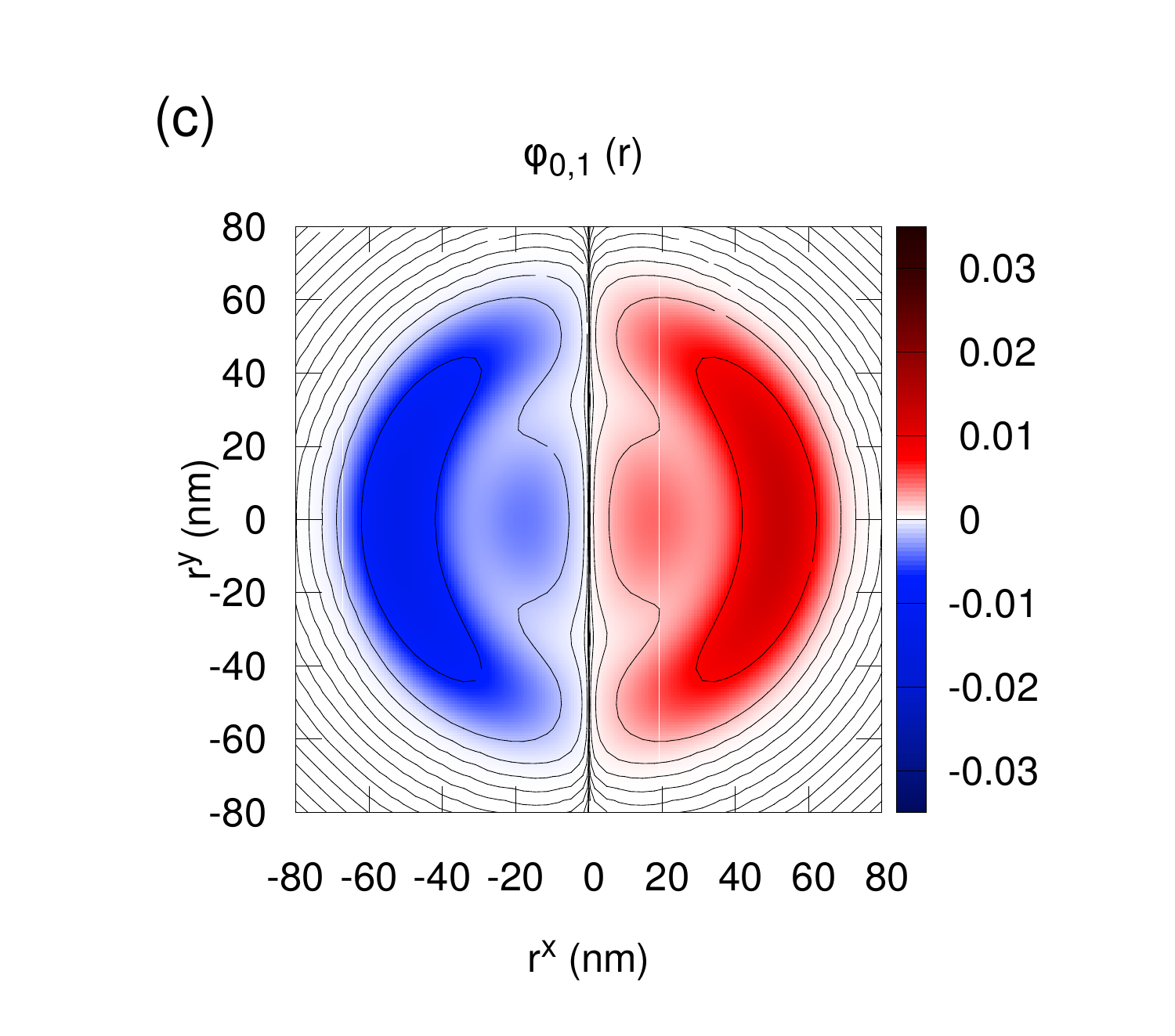} \\
 \includegraphics[width=.44\linewidth]{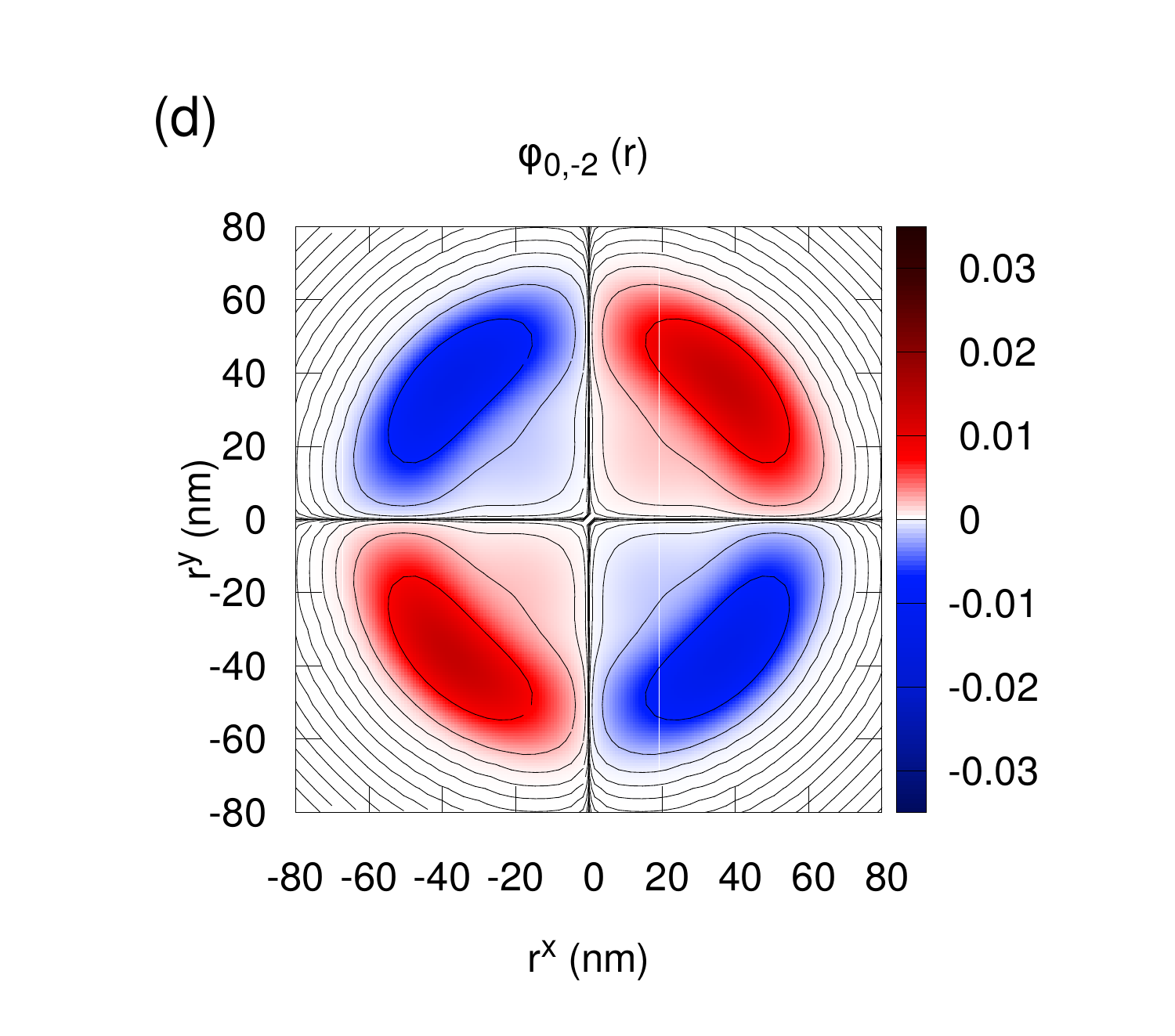}
 \includegraphics[width=.44\linewidth]{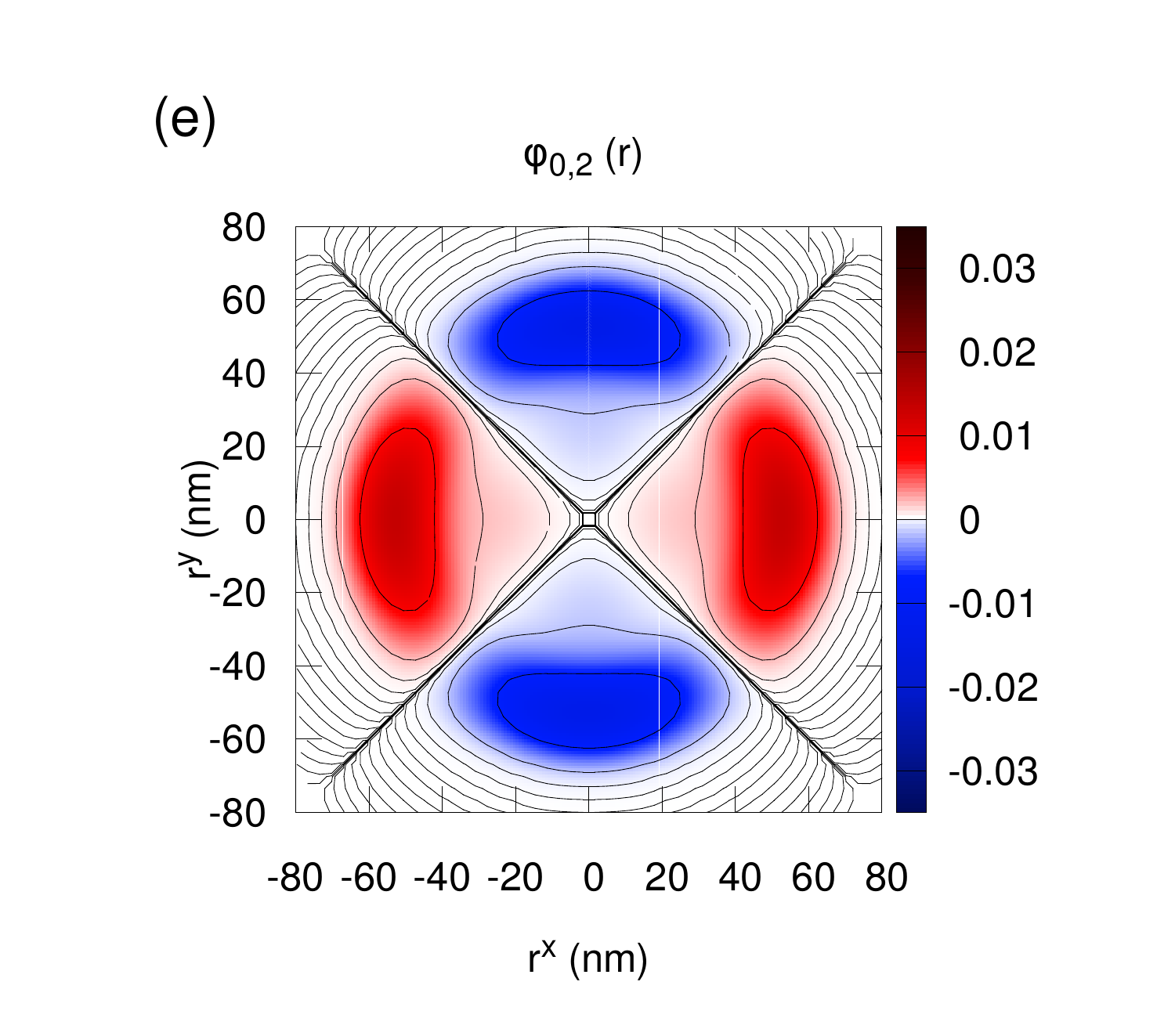}
 \caption{Starting single-particle wavefunctions for $n=0$, $l=0,\mp1,\mp2$ ((a) -- (e)), respectively, all for the quantum dot potential $V_\mathrm{QD}=0$,
 taken to define the field operator in Eq. \eqref{eq:field_operator}.
 Note their similarity to the $s$, $p_x$, $p_y$, $d_{xy}$ and $d_{x^2-y^2}$ atomic states, respectively.}
  \label{Fig:R0}
\end{figure}
\begin{figure}[p]
 \includegraphics[width=.44\linewidth]{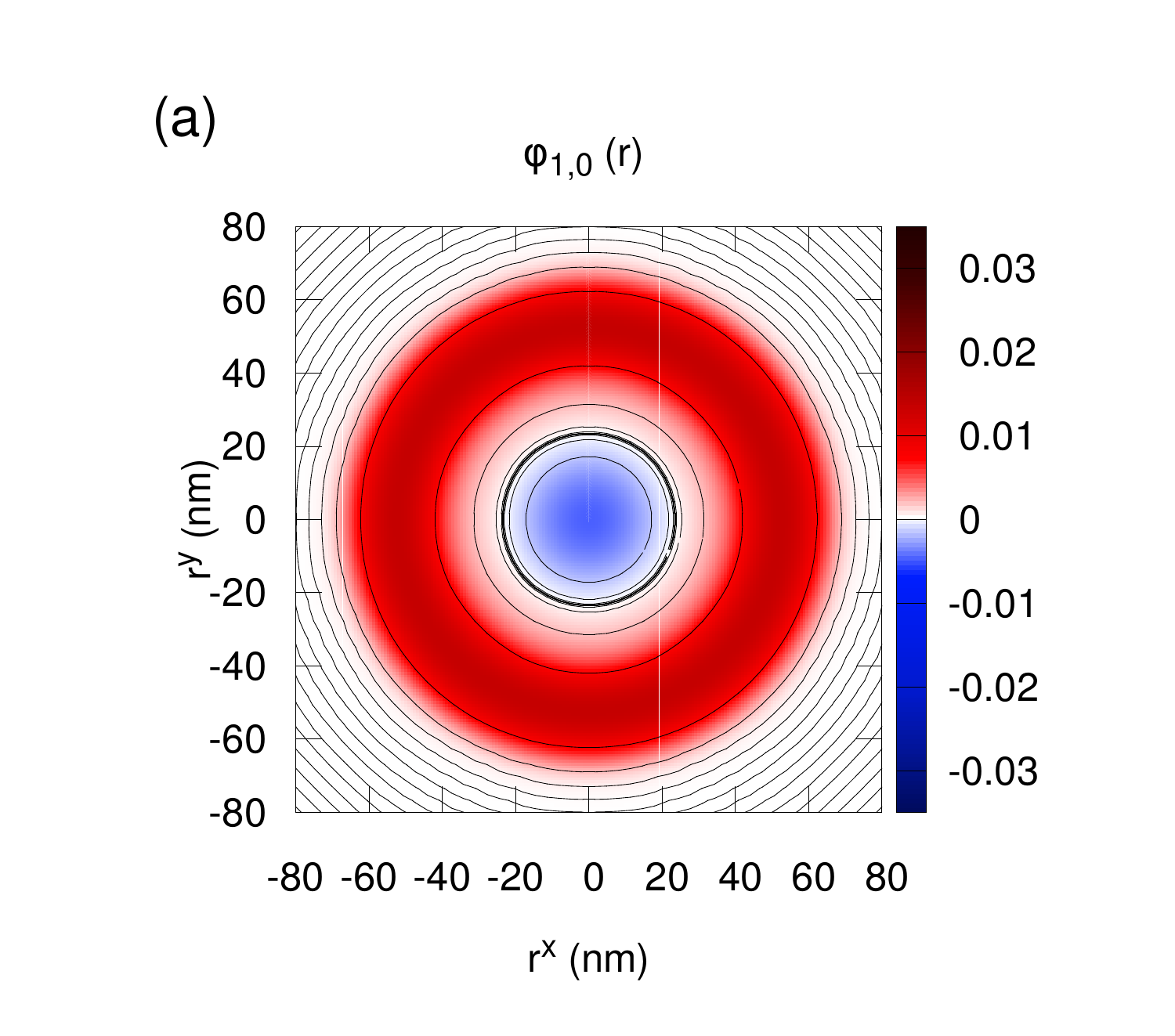} \\
 \includegraphics[width=.44\linewidth]{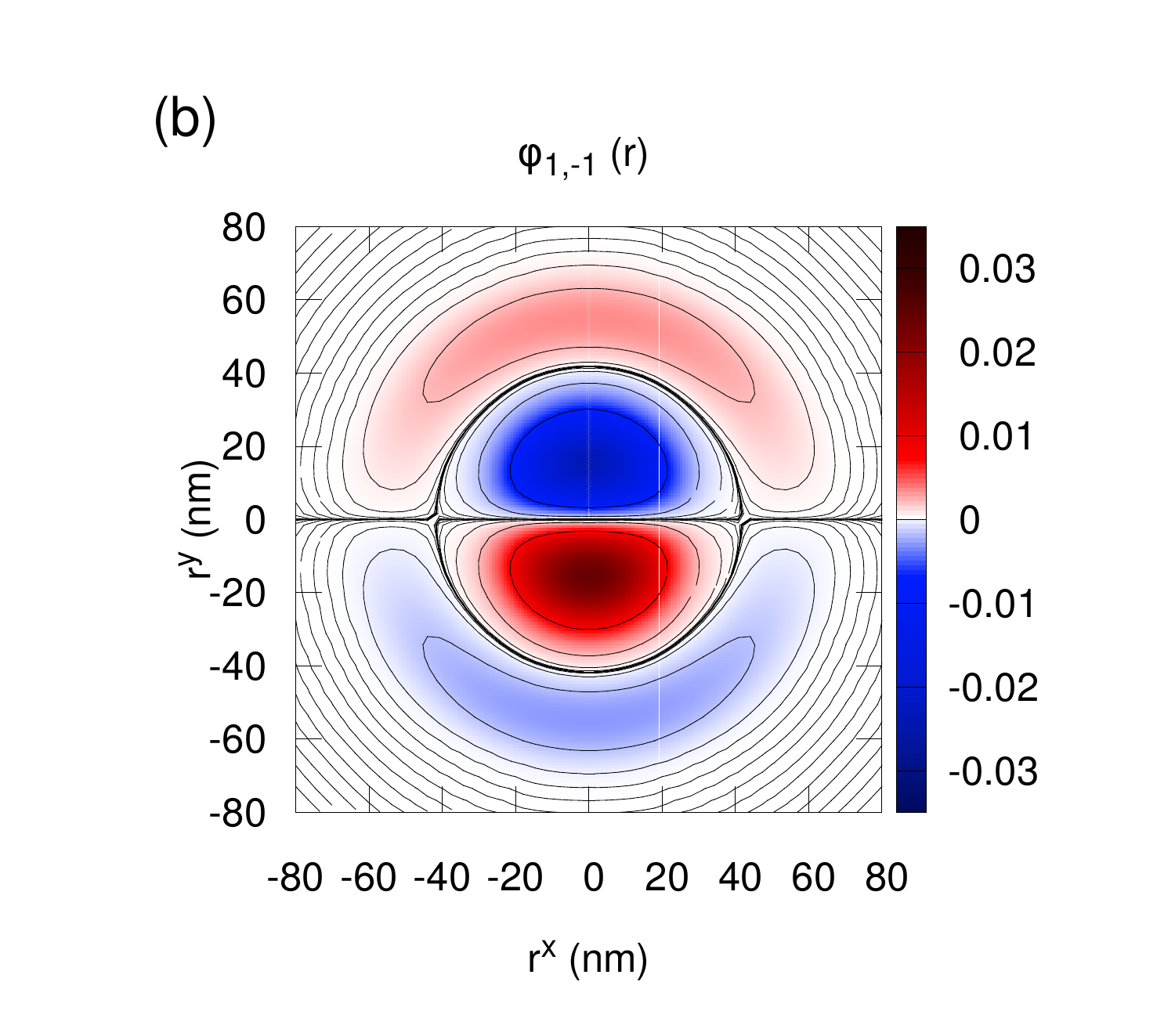}
 \includegraphics[width=.44\linewidth]{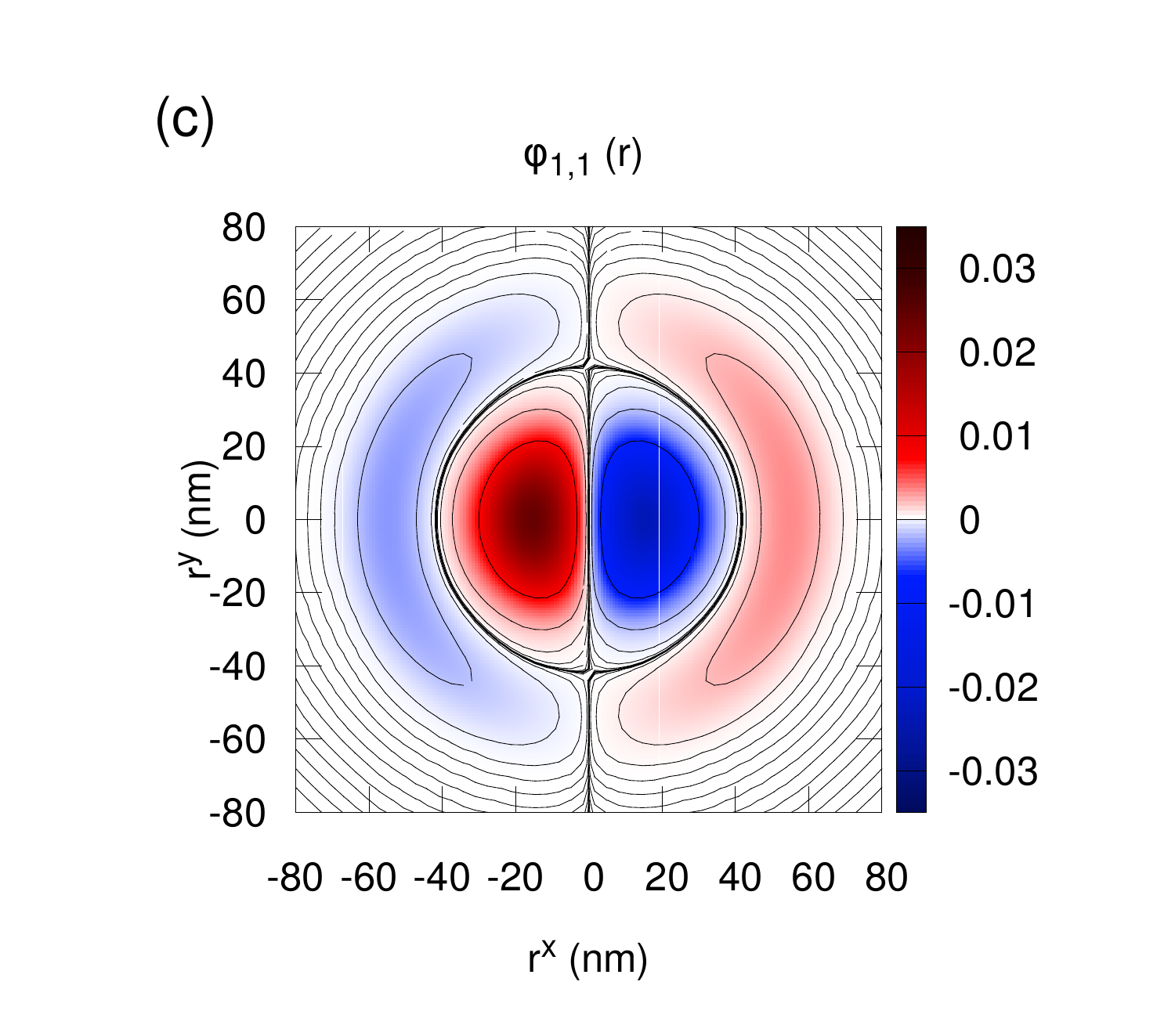} \\
 \includegraphics[width=.44\linewidth]{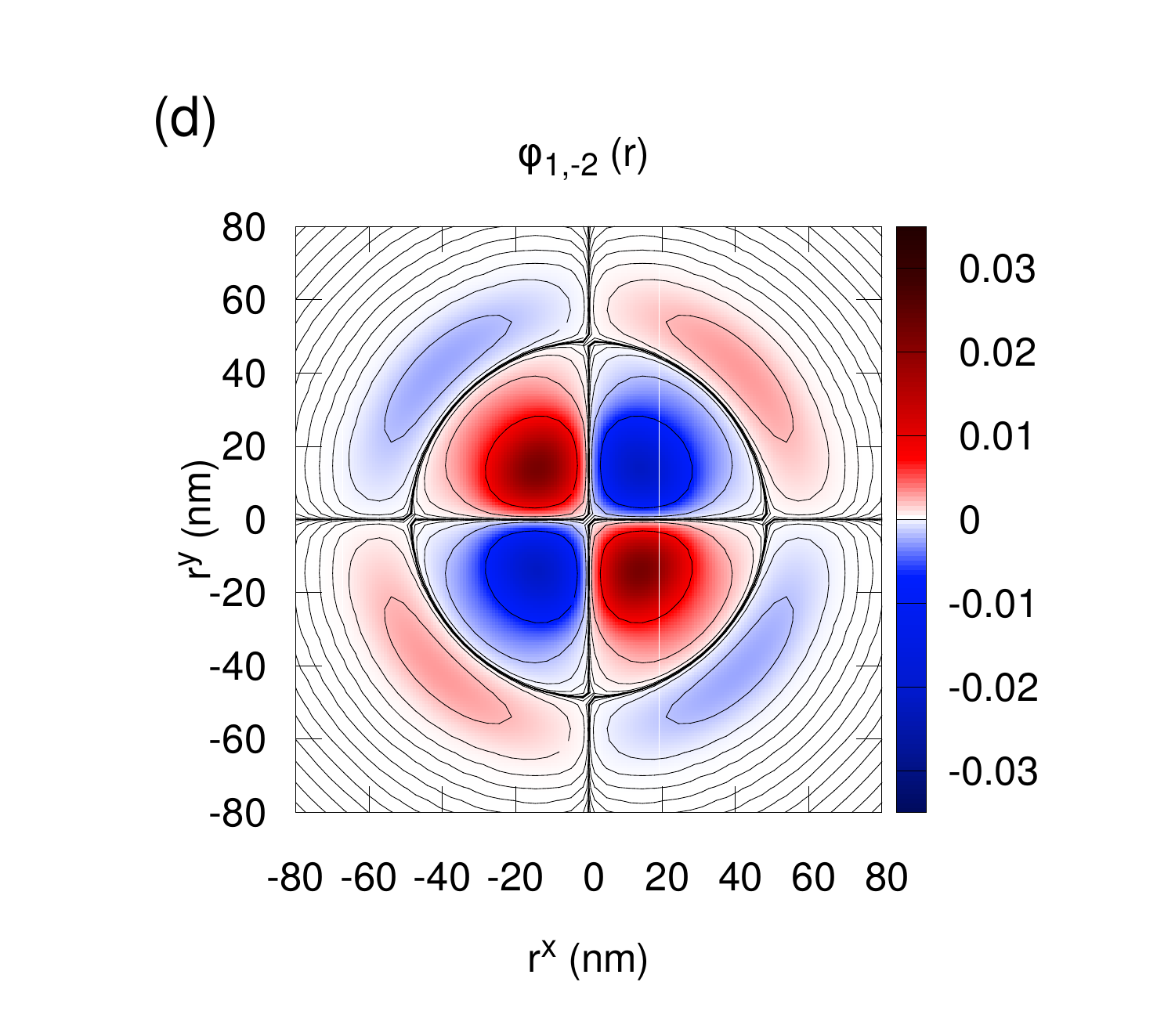} 
 \includegraphics[width=.44\linewidth]{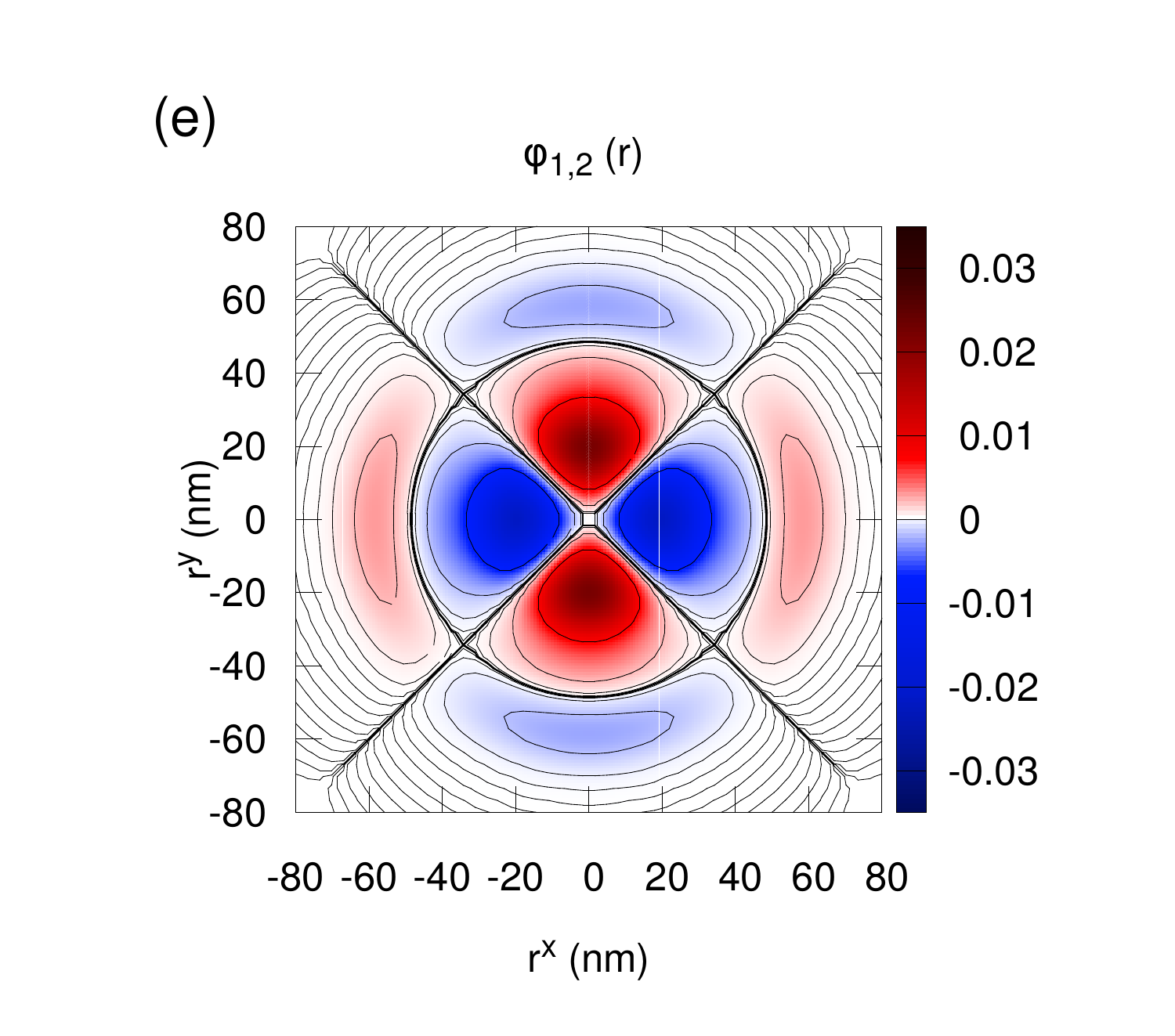}
 \caption{Starting single-particle wavefunctions for $n=1$, $l=0,\mp1,\mp2$ ((a) -- (e)), respectively, all for the quantum dot potential $V_\mathrm{QD}=0$,
 taken together with those depicted in Fig.~\ref{Fig:R0} to define the field operator \eqref{eq:field_operator}.
 The subsidiary (external) maxima and minima reflect the part associated with the presence of the ring.}
  \label{Fig:R1}
\end{figure}

\newpage
\section{Detailed values of microscopic interaction parameters in Fock space}

For the sake of completeness we provide the detailed numerical values of selected Coulomb interaction parameters as a function
of the relative QD potential $V_\mathrm{QD}$. Note that we have included also the 3- and 4-state interaction
parameters, often ignored in many-particle considerations. Although the values of those last parameters are small, they are important 
as the corresponding number of such four-state terms (c.f. ~\ref{tab:V}) is the largest and equal to $5040$ (when we disregard symmetries leading to their degeneracy).  

\begin{table*}[!htbp]
\caption{All Hubbard intrastate repulsion amplitudes $U_{i} \equiv V_{iiii}$ (in meV) for different QD potentials.}
\label{tab:U}
\begin{tabular}{l||r|r|r|r|r|r|r|r|r|r|r|r|r}
$V_\mathrm{QD}$ (meV)	&	-6	&	-5	&	-4	&	-3	&	-2	&	-1	&	0	&	1	&	2	&	3	&	4	&	5	&	6\\\hline\hline
$U_{(0 \ 0)}$	&	8.67 & 	8.60 & 	8.59 & 	8.51 & 	8.55 & 	8.38 & 	8.33 & 	8.08 & 	7.15 & 	2.97 & 	2.90 & 	2.88 & 	2.92  \\\hline
$U_{(0 \ 1)}$	&	7.61 & 	7.49 & 	7.20 & 	6.27 & 	3.67 & 	3.32 & 	3.41 & 	3.44 & 	3.45 & 	3.50 & 	3.48 & 	3.51 & 	3.49  \\\hline
$U_{(0 \ \bar{1})}$	&	7.63 & 	7.47 & 	7.14 & 	6.21 & 	3.63 & 	3.29 & 	3.39 & 	3.47 & 	3.43 & 	3.49 & 	3.49 & 	3.51 & 	3.53  \\\hline
$U_{(0 \ 2)}$	&	3.19 & 	3.23 & 	3.28 & 	3.29 & 	3.28 & 	3.29 & 	3.28 & 	3.32 & 	3.28 & 	3.30 & 	3.27 & 	3.27 & 	3.26  \\\hline
$U_{(0 \ \bar{2})}$	&	3.26 & 	3.24 & 	3.25 & 	3.27 & 	3.31 & 	3.26 & 	3.27 & 	3.30 & 	3.26 & 	3.32 & 	3.26 & 	3.29 & 	3.26  \\\hline
$U_{(1 \ 0)}$	&	2.94 & 	2.90 & 	2.89 & 	2.93 & 	2.91 & 	2.89 & 	2.89 & 	2.89 & 	2.87 & 	5.60 & 	7.74 & 	7.97 & 	7.96  \\\hline
$U_{(1 \ 1)}$	&	3.47 & 	3.51 & 	3.40 & 	3.33 & 	4.25 & 	6.50 & 	7.02 & 	7.11 & 	7.15 & 	7.01 & 	6.89 & 	6.83 & 	6.64  \\\hline
$U_{(1 \ \bar{1})}$	&	3.46 & 	3.46 & 	3.42 & 	3.28 & 	4.27 & 	6.50 & 	6.97 & 	7.12 & 	7.08 & 	7.02 & 	6.88 & 	6.83 & 	6.66  \\\hline
$U_{(1 \ 2)}$	&	6.18 & 	6.31 & 	6.27 & 	6.26 & 	6.02 & 	5.93 & 	5.86 & 	5.76 & 	5.61 & 	5.35 & 	5.18 & 	4.74 & 	4.36  \\\hline
$U_{(1 \ \bar{2})}$	&	6.20 & 	6.29 & 	6.30 & 	6.25 & 	6.15 & 	6.04 & 	5.93 & 	5.82 & 	5.63 & 	5.29 & 	5.09 & 	4.75 & 	4.40  
 \end{tabular} 
\end{table*}

\begin{table*}[!htbp]
\caption{Selected values of the interstate Coulomb repulsion amplitudes $K_{ij} \equiv V_{ijij}$ (in meV) for different QD potentials.}
\label{tab:K}
\begin{tabular}{l||r|r|r|r|r|r|r|r|r|r|r|r|r}
$V_\mathrm{QD}$ (meV)	&	-6	&	-5	&	-4	&	-3	&	-2	&	-1	&	0	&	1	&	2	&	3	&	4	&	5	&	6\\\hline\hline
$K_{(0 \ 0),(0 \ 1)}$	&	7.20 & 	7.15 & 	6.99 & 	6.46 & 	4.26 & 	2.67 & 	2.38 & 	2.29 & 	2.32 & 	2.73 & 	2.90 & 	2.84 & 	2.93  \\\hline
$K_{(0 \ 0),(0 \ 2)}$	&	2.38 & 	2.28 & 	2.25 & 	2.26 & 	2.23 & 	2.22 & 	2.23 & 	2.22 & 	2.30 & 	2.74 & 	2.88 & 	2.91 & 	2.91  \\\hline
$K_{(0 \ 0),(1 \ 0)}$	&	2.28 & 	2.27 & 	2.27 & 	2.27 & 	2.27 & 	2.27 & 	2.29 & 	2.38 & 	2.88 & 	3.53 & 	2.45 & 	2.29 & 	2.26  \\\hline
$K_{(0 \ 0),(1 \ 1)}$	&	2.28 & 	2.31 & 	2.43 & 	2.90 & 	5.03 & 	6.57 & 	6.81 & 	6.72 & 	6.37 & 	3.47 & 	2.44 & 	2.34 & 	2.31  \\\hline
$K_{(0 \ 0),(1 \ 2)}$	&	6.17 & 	6.15 & 	6.11 & 	6.11 & 	6.03 & 	5.92 & 	5.87 & 	5.66 & 	5.35 & 	3.17 & 	2.50 & 	2.44 & 	2.46  \\\hline
$K_{(0 \ 1),(0 \ \bar{1})}$	&	5.89 & 	5.84 & 	5.60 & 	4.90 & 	2.89 & 	2.35 & 	2.32 & 	2.34 & 	2.36 & 	2.35 & 	2.36 & 	2.33 & 	2.37  \\\hline
$K_{(0 \ 2),(0 \ \bar{2})}$	&	2.54 & 	2.54 & 	2.53 & 	2.53 & 	2.57 & 	2.55 & 	2.54 & 	2.55 & 	2.54 & 	2.53 & 	2.53 & 	2.52 & 	2.55  \\\hline
$K_{(1 \ 1),(1 \ \bar{1})}$	&	2.33 & 	2.32 & 	2.35 & 	2.37 & 	3.41 & 	5.11 & 	5.47 & 	5.51 & 	5.47 & 	5.47 & 	5.35 & 	5.29 & 	5.14  
 \end{tabular} 
\end{table*}

\begin{table*}[!htbp]
\caption{Selected values of the interstate exchange integral $J_{ij} \equiv V_{ijji}$ (in meV) for different QD potentials.}
\label{tab:J}
\begin{tabular}{l||r|r|r|r|r|r|r|r|r|r|r|r|r}
$V_\mathrm{QD}$ (meV)	&	-6	&	-5	&	-4	&	-3	&	-2	&	-1	&	0	&	1	&	2	&	3	&	4	&	5	&	6\\\hline\hline
$J_{(0 \ 0),(0 \ 1)}$	&	2.40 & 	2.37 & 	2.31 & 	2.03 & 	1.00 & 	0.24 & 	0.10 & 	0.08 & 	0.20 & 	1.20 & 	1.52 & 	1.53 & 	1.55  \\\hline
$J_{(0 \ 0),(0 \ 2)}$	&	0.04 & 	0.02 & 	0.02 & 	0.01 & 	0.01 & 	0.01 & 	0.01 & 	0.03 & 	0.12 & 	0.86 & 	1.05 & 	1.15 & 	1.14  \\\hline
$J_{(0 \ 0),(1 \ 0)}$	&	0.04 & 	0.03 & 	0.03 & 	0.03 & 	0.04 & 	0.05 & 	0.08 & 	0.17 & 	0.62 & 	1.28 & 	0.25 & 	0.09 & 	0.05  \\\hline
$J_{(0 \ 0),(1 \ 1)}$	&	0.04 & 	0.06 & 	0.11 & 	0.35 & 	1.39 & 	2.16 & 	2.26 & 	2.20 & 	2.09 & 	0.62 & 	0.11 & 	0.06 & 	0.04  \\\hline
$J_{(0 \ 0),(1 \ 2)}$	&	1.09 & 	1.06 & 	1.06 & 	1.06 & 	1.04 & 	0.98 & 	0.97 & 	0.94 & 	0.88 & 	0.30 & 	0.11 & 	0.10 & 	0.10  \\\hline
$J_{(0 \ 1),(0 \ \bar{1})}$	&	0.88 & 	0.87 & 	0.76 & 	0.66 & 	0.38 & 	0.48 & 	0.53 & 	0.55 & 	0.57 & 	0.56 & 	0.57 & 	0.58 & 	0.56  \\\hline
$J_{(0 \ 2),(0 \ \bar{2})}$	&	0.34 & 	0.36 & 	0.37 & 	0.36 & 	0.39 & 	0.37 & 	0.35 & 	0.40 & 	0.36 & 	0.38 & 	0.38 & 	0.39 & 	0.39  \\\hline
$J_{(1 \ 1),(1 \ \bar{1})}$	&	0.57 & 	0.57 & 	0.53 & 	0.47 & 	0.43 & 	0.71 & 	0.80 & 	0.79 & 	0.80 & 	0.79 & 	0.77 & 	0.77 & 	0.73  
 \end{tabular} 
\end{table*}

\begin{table*}[!htbp]
\caption{Selected correlated interstate hopping parameter values $C_{ij} \equiv V_{ijjj}$ (in meV) for different QD potentials.}
\label{tab:C}
\begin{tabular}{l||r|r|r|r|r|r|r|r|r|r|r|r|r}
$V_\mathrm{QD}$ (meV)	&	-6	&	-5	&	-4	&	-3	&	-2	&	-1	&	0	&	1	&	2	&	3	&	4	&	5	&	6\\\hline\hline
$C_{(0 \ 0),(0 \ 1)}$	&	0.05 & 	-0.03 & 	0.01 & 	-0.06 & 	-0.01 & 	-0.01 & 	0.01 & 	0 & 	0 & 	-0.02 & 	0 & 	-0.01 & 	0.03  \\\hline
$C_{(0 \ 0),(0 \ 2)}$	&	0 & 	0.01 & 	0 & 	0 & 	0 & 	0 & 	0 & 	0 & 	-0.01 & 	0.02 & 	0.02 & 	0.01 & 	0.01  \\\hline
$C_{(0 \ 0),(1 \ 0)}$	&	-0.38 & 	-0.37 & 	-0.37 & 	-0.39 & 	-0.44 & 	-0.51 & 	-0.65 & 	-0.94 & 	-1.65 & 	-0.47 & 	0.09 & 	0.08 & 	0.06  \\\hline
$C_{(0 \ 0),(1 \ 1)}$	&	0 & 	-0.01 & 	0 & 	0 & 	-0.05 & 	0.02 & 	-0.02 & 	0 & 	0.03 & 	0 & 	0 & 	0 & 	0  \\\hline
$C_{(0 \ 0),(1 \ 2)}$	&	0.03 & 	0 & 	-0.03 & 	0.02 & 	-0.03 & 	-0.01 & 	-0.01 & 	0.02 & 	0 & 	-0.02 & 	0 & 	0.01 & 	-0.01  \\\hline
$C_{(0 \ 1),(0 \ \bar{1})}$	&	0.03 & 	0.01 & 	0.01 & 	-0.01 & 	-0.01 & 	-0.02 & 	0 & 	0 & 	-0.01 & 	0 & 	-0.01 & 	0 & 	-0.01  \\\hline
$C_{(0 \ 2),(0 \ \bar{2})}$	&	-0.03 & 	0.03 & 	-0.01 & 	0 & 	0 & 	0.02 & 	-0.01 & 	-0.01 & 	0 & 	0.03 & 	-0.01 & 	-0.02 & 	-0.01  \\\hline
$C_{(1 \ 1),(1 \ \bar{1})}$	&	0 & 	0.02 & 	0 & 	0 & 	0.01 & 	-0.03 & 	-0.02 & 	0.04 & 	0 & 	0.02 & 	-0.02 & 	0 & 	-0.01  
 \end{tabular} 
\end{table*}

\begin{table*}[!htbp]
\caption{Selected three- and four-state parameters $V_{[ijkl]}$ (in meV) for different QD potentials.}
\label{tab:V}
\begin{tabular}{l||r|r|r|r|r|r|r|r|r|r|r|r|r}
$V_\mathrm{QD}$ (meV)	&	-6	&	-5	&	-4	&	-3	&	-2	&	-1	&	0	&	1	&	2	&	3	&	4	&	5	&	6\\\hline\hline
$V_{(0 \ 0),(0 \ 1),(0 \ 1),(0 \ 2)}$	&	0.34 & 	0.27 & 	0.23 & 	0.23 & 	0.21 & 	0.15 & 	0.14 & 	0.19 & 	0.37 & 	0.97 & 	1.08 & 	1.10 & 	1.12  \\\hline
$V_{(0 \ 0),(0 \ 1),(0 \ \bar{2}),(0 \ \bar{1})}$	&	0.19 & 	0.14 & 	0.11 & 	0.08 & 	0.06 & 	0.06 & 	0.07 & 	0.12 & 	0.25 & 	0.68 & 	0.80 & 	0.81 & 	0.82  \\\hline
$V_{(0 \ 0),(0 \ \bar{1}),(0 \ 1),(0 \ \bar{2})}$	&	0.35 & 	0.27 & 	0.23 & 	0.23 & 	0.21 & 	0.15 & 	0.15 & 	0.19 & 	0.37 & 	0.95 & 	1.06 & 	1.11 & 	1.09  \\\hline
$V_{(0 \ 0),(0 \ 2),(0 \ 2),(0 \ \bar{1})}$	&	0 & 	0 & 	0 & 	0 & 	0 & 	0 & 	0 & 	0 & 	0 & 	0.03 & 	0.04 & 	-0.01 & 	0.01  \\\hline
$V_{(0 \ 1),(0 \ 1),(1 \ 0),(0 \ 0)}$	&	0.04 & 	0.03 & 	0.04 & 	0.09 & 	0.18 & 	0.18 & 	0.20 & 	0.26 & 	0.48 & 	0.67 & 	0.29 & 	0.18 & 	0.12  \\\hline
$V_{(1 \ 0),(0 \ 2),(1 \ \bar{1}),(0 \ \bar{1})}$	&	-0.16 & 	-0.19 & 	-0.24 & 	-0.40 & 	-0.57 & 	-0.33 & 	-0.23 & 	-0.19 & 	-0.19 & 	-0.28 & 	-0.27 & 	-0.27 & 	-0.28  
 \end{tabular} 
\end{table*}

\newpage
$ $
\newpage
\section{DRN in the correlated state: 2- and 3-electrons particle density}

Here we provide the characteristics of the first excited state for $N_e=2$ for two additional values of $V_\mathrm{QD}$.
\begin{figure*}[h!]
  \includegraphics[width=0.9\textwidth]{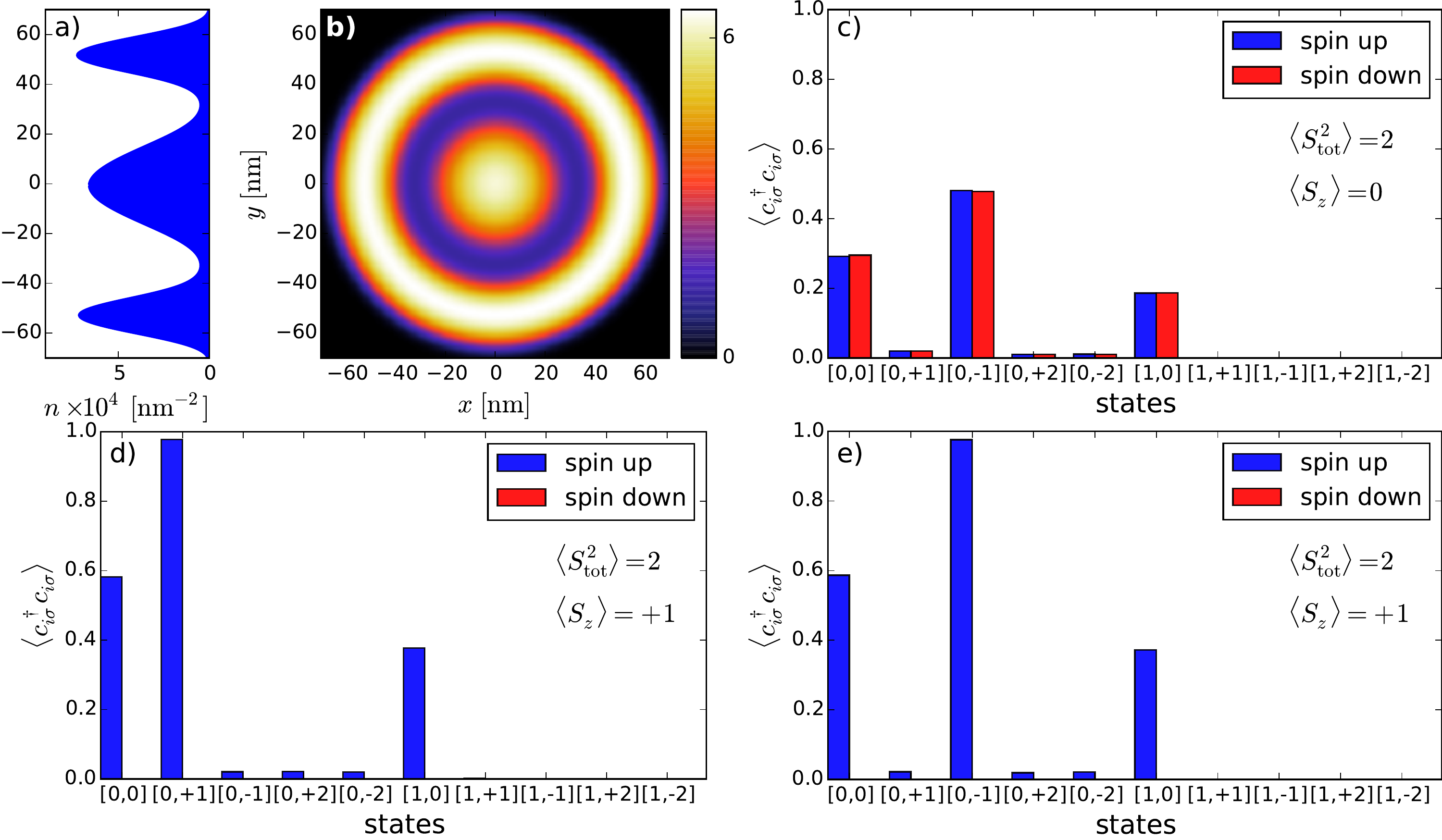}
\caption{The same as in Fig. \ref{Fig:2DRN_m4p4_exc}, but for $V_\mathrm{QD}=2$ meV.}
  \label{Fig:Supp1}
\end{figure*}

\begin{figure*}[h!]
  \includegraphics[width=0.9\textwidth]{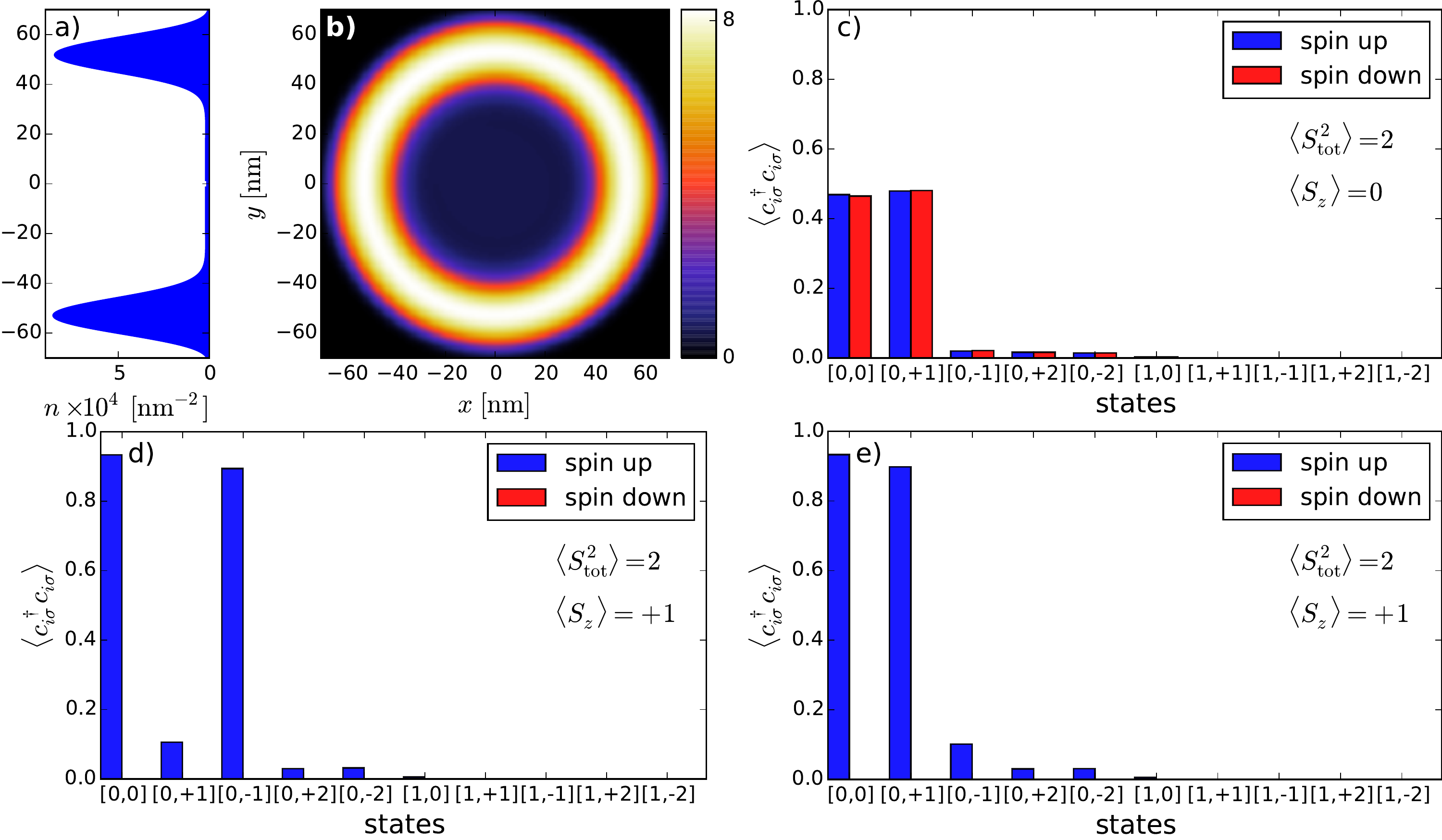}
\caption{The same as in Fig. \ref{Fig:2DRN_m4p4_exc}, but for $V_\mathrm{QD}=4$ meV.}
    \label{Fig:Supp2}
\end{figure*}

\newpage
\section{The degree of degeneracy in the ground- and first-excited state}
 In this Supplement we list the degeneracies of the multiparticle states. The first factor of the total degeneracy is due to the $S^z_{tot}$ ($2S^z_{tot}+1$ values). The additional degeneracy represents an emergent chirality and is related to the number of ways the single-electron current can compose the total orbital current.
 
\begin{table*}[!htbp]
\caption{The degrees of degeneracy for different QD potentials, with $N_e=2$, $3$.}
\label{tab:degeneracy}
\begin{tabular}{r||c|c|c|c||c|c|c|c||}
 & \mc{4}{c||}{2 electrons} & \mc{4}{c||}{3 electrons} \\
 & \mc{2}{c|}{ground state} & \mc{2}{c||}{first-excited state}& \mc{2}{c|}{ground state} & \mc{2}{c||}{first-excited state} \\\hline
 $V_\mathrm{QD}$ (meV) & deg. & $S_{tot}$ & deg. & $S_{tot}$ & deg. & $S_{tot}$ & deg. & $S_{tot}$ \\\hline\hline
-6 & 1 & 0 & $3 \times 2$ & 1 & $2 \times 3$ & ${1}/{2}$ & $2 \times 2$ & ${1}/{2}$ \\\hline
-5 & 1 & 0 & $3 \times 2$ & 1 & $2 \times 3$ & ${1}/{2}$ & $2 \times 2$ & ${1}/{2}$ \\\hline
-4 & 1 & 0 & 3 & 1 & $2 \times 3$ & ${1}/{2}$ & $4 \times 2$ & ${3}/{2}$ \\\hline
-3 & 1 & 0 & 3 & 1 & $2 \times 3$ & ${1}/{2}$ & $4 \times 2$ & ${3}/{2}$ \\\hline
-2 & 1 & 0 & 3 & 1 & $2 \times 3$ & ${1}/{2}$ & $4 \times 2$ & ${3}/{2}$ \\\hline
-1 & 1 & 0 & 3 & 1 & $2 \times 3$ & ${1}/{2}$ & $4 \times 2$ & ${3}/{2}$ \\\hline
0 & 1 & 0 & 3 & 1 & $2 \times 3$ & ${1}/{2}$ & $4 \times 2$ & ${3}/{2}$ \\\hline
1 & 1 & 0 & 3 & 1 & $2 \times 3$ & ${1}/{2}$ & $4 \times 2$ & ${3}/{2}$ \\\hline
2 & 1 & 0 & $3 \times 2$ & 1 & 4 & ${3}/{2}$ & $2 \times 2$ & ${1}/{2}$ \\\hline
3 & 1 & 0 & $3 \times 2$ & 1 & 4 & ${3}/{2}$ & $2 \times 2$ & ${1}/{2}$ \\\hline
4 & 1 & 0 & $3 \times 2$ & 1 & 4 & ${3}/{2}$ & $2 \times 2$ & ${1}/{2}$ \\\hline
5 & 1 & 0 & $3 \times 2$ & 1 & 4 & ${3}/{2}$ & $2 \times 2$ & ${1}/{2}$ \\\hline
6 & 1 & 0 & $3 \times 2$ & 1 & 4 & ${3}/{2}$ & $2 \times 2$ & ${1}/{2}$
\end{tabular} 
\end{table*}

\section{Average total momentum along the z-axis}

We define the average total momentum along the $z$-axis as
\begin{align}
    \label{eq:av_tot_mom}
    \average{L^z} \equiv  \sum_\sigma \matrixel{\Phi}{\int d^3 r \hat{\Psi}_{\sigma}^{\dagger}(\vec{r}) \hat{L}^z \hat{\Psi}_{\sigma}^{\phantom{\dagger}}(\vec{r})}{\Phi} = \sum_{\sigma, i, j} \int d^3 r \varphi_{i\sigma}^*  \hat{L}^z \varphi_{j\sigma} \average{\CR{i}{2} \AN{j}{2}} ,
\end{align}
where $\ket{\Phi}$ is the state in the Fock space, $\hat{\Psi}_{\sigma}^{\phantom{\dagger}}(\vec{r})$ is the field operator \eqref{eq:field_operator}, $\{ \varphi_i \}$ is the single-particle wave-function basis, $i$ and $j$ are the pairs of quantum numbers $[n,l]$ defining the single-particle state, and
\begin{align}
\label{eq:momentum_z}
\hat{L}^z \equiv - i \hbar \partial_\phi    
\end{align}
is the single-particle $z$-component momentum operator. To calculate \eqref{eq:av_tot_mom} we must calculate how \eqref{eq:momentum_z} influences the single-particle wave-function basis $\{ \varphi_i \}$. First, let us consider \eqref{eq:momentum_z} acting on the simpler single-particle wave-function basis $\{ \psi_i \}$
\begin{align}
    \label{eq:mom_on_psi}
    \hat{L}^z \psi_{nl} (\vec{r}) \equiv - i \hbar \partial_\phi \psi_{nl} (\vec{r}) =  - i \hbar \partial_\phi R_{nl} (r) e^{i l \phi} = \hbar l \psi_{nl} (\vec{r}).
\end{align}
Now, $\hat{L}^z$ acting $\{ \varphi_i \}$ comes down to three cases
\begin{subequations}
\begin{align}
    \label{eq:case_l=0}
    &l = 0: \ \ \ - i \hbar \partial_\phi \varphi_{n0} (\vec{r}) = - i \hbar \partial_\phi \psi_{n0} (\vec{r}) = 0, \\
    \label{eq:case_l>0}
    &l > 0: \ \ \ - i \hbar \partial_\phi \varphi_{nl} (\vec{r}) = - i \hbar \partial_\phi \frac{\psi_{nl} (\vec{r}) + \psi_{n\bar{l}} (\vec{r})}{\sqrt{2}} = i \hbar l \frac{\psi_{nl} (\vec{r}) - \psi_{n\bar{l}} (\vec{r})}{\sqrt{2}i} = i \hbar l \varphi_{n\bar{l}} (\vec{r}), \\
    \label{eq:case_l<0}
    &l < 0: \ \ \ - i \hbar \partial_\phi \varphi_{nl} (\vec{r}) = - i \hbar \partial_\phi \frac{\psi_{n\bar{l}} (\vec{r}) - \psi_{nl} (\vec{r})}{\sqrt{2} i } = i \hbar l \frac{\psi_{n\bar{l}} (\vec{r}) + \psi_{nl} (\vec{r})}{\sqrt{2}} = i \hbar l \varphi_{n\bar{l}} (\vec{r}).
\end{align}
\end{subequations}
Eventually, due to the orthogonality of basis $\{ \varphi_i \}$ the only non-zero elements in the sum \eqref{eq:av_tot_mom} are
\begin{align}
    \label{eq:av_tot_mom_ortho}
    \average{L^z} = \sum_{\sigma, n, l>0} i \hbar l \left( \average{\CR{n \bar{l}}{0} \AN{n l }{0}} - \average{\CR{n l}{0} \AN{n \bar{l}}{0}} \right).
\end{align}
It means that as long as the symmetry condition $ \average{\CR{n \bar{l}}{0} \AN{n l }{0}} = \average{\CR{n l}{0} \AN{n \bar{l}}{0}}$ is fulfilled (which is the case here), the average total momentum along $z$-axis will be equal to zero.
\label{PAG:endSupp}

\end{document}